\documentclass[manuscript]{acmart} 
\AtBeginDocument{%
  }

\usepackage{comment}
\usepackage{tgtermes}
\usepackage[utf8]{inputenc}
\usepackage{amsmath,amsfonts}
\usepackage{graphicx}
\usepackage{amsthm}
\usepackage{xcolor}
\usepackage{indentfirst}
\usepackage{multicol}
\usepackage{multirow}
\usepackage{threeparttable}
\usepackage{algorithmicx}
\usepackage{algpseudocode}
\usepackage{wrapfig}
\usepackage{graphicx}
\usepackage{tabularx}
\usepackage{subcaption}
\usepackage{tabularray}   
\usepackage{booktabs}
\usepackage{tikz}
\usetikzlibrary{matrix}
\usepackage{verbatim}
\usepackage{pgfplots}
\usepackage{pgfplotstable}
\usetikzlibrary{pgfplots.groupplots}
\usepackage{pgf-pie}
\usepackage{algorithm}
\usepackage{balance}
\usepackage{enumitem}
\usepackage{soul}
\usepackage{listings}
\usepackage{xcolor}
\usepackage[labelformat=simple]{subcaption}

\begin{document}

\title{FixV2W: Correcting Invalid CVE-CWE Mappings with Knowledge Graph Embeddings}

\author{Şevval Şimşek}
\affiliation{%
  \institution{Boston University}
  \city{Boston}
  \country{USA}}
\email{sevvals@bu.edu}

\author{Varsha Athreya}
\affiliation{%
  \institution{Boston University}
  \city{Boston}
  \country{USA}}
\email{vathreya@bu.edu}

\author{David Starobinski}
\affiliation{%
  \institution{Boston University}
  \city{Boston}
  \country{USA}}
\email{staro@bu.edu}

\renewcommand{\shortauthors}{Şimşek \emph{et al.}}
\newcommand{\fixvw}{\textit{FixV2W}}

\begin{abstract}
Accurate mapping between Common Vulnerabilities and Exposures (CVE) and Common Weakness Enumeration (CWE) entries is critical for effective vulnerability management and risk assessment. However, public databases, such as the National Vulnerability Database (NVD), suffer from inconsistent and incomplete CVE–CWE mappings, complicating automated analysis and remediation. We introduce FixV2W, a lightweight approach that leverages knowledge graph embeddings and longitudinal trends to improve mapping accuracy of the NVD.
FixV2W systematically analyzes historical remapping patterns and leverages hierarchical relationships within NVD and CWE data to predict more precise CWE mappings for vulnerabilities linked to Prohibited or Discouraged categories. We run extensive experimental evaluation of FixV2W, based on test data set collected between August 2021 and December 2024. Considering the Top-10 ranked predictions, the results show that FixV2W predicts the correct CWE mappings for 69\% of exploited vulnerabilities that had invalid CWEs before they were exploited. We also show that  FixV2W significantly improves the performance of ML models relying on NVD data. For instance, for a model geared at uncovering unknown CVE-CWE mappings, FixV2W  improves the Mean Reciprocal Rank (MRR) from 0.174 to 0.608.  
These results show that FixV2W is a promising approach to identify and thwart emerging threats.

\end{abstract}



\maketitle

\section{Introduction}


 
Threat databases, such as Snyk Vulnerability Database~\cite{snyk-vdb}, Open Source Vulnerabilities (OSV) Database~\cite{osv}, and others~\cite{redhat-vdb,mend-db}, play a critical role in the software security ecosystem by providing structured, actionable information about known vulnerabilities, threats and mitigation strategies~\cite{LI2023111679}. 
Collectively, these databases support vulnerability management, threat detection, risk prioritization, and the development of more secure software. 

While there exist many commercial and proprietary vulnerability databases, the National Vulnerability Database (NVD) remains the central source of vulnerability information that many tools and other databases ~\cite{snyk-vdb, osv, redhat-vdb, mend-db} rely on. The public availability and widespread usage of the NVD make it a foundational component of cybersecurity practices, offering standardized data formats, severity metrics, and consistent identifiers. This ensures interoperability among security tools, facilitates vulnerability management workflows, and promotes transparency across the cybersecurity ecosystem. 

The NVD~\cite{nvd} serves as a key resource by providing  additional information (called \emph{enrichment metadata}) for Common Vulnerabilities and Exposures (CVEs)~\cite{cve}, including mappings to Common Weakness Enumeration (CWE)~\cite{cwe}, Common Platform Enumeration (CPE)~\cite{cpe} and CVSS (Common Vulnerability Scoring System)~\cite{first_cvss} scores. This structured metadata allows automated tools and security teams to assess and prioritize vulnerabilities effectively. In particular, CWE mappings provide critical insight into the \emph{root causes} of vulnerabilities, allowing developers and analysts to better understand and prevent recurring classes of security issues~\cite{shostack2014threat}. Understanding and resolving a software flaw related to a weakness helps resolving a host of CVEs that emerges from this flaw.
Early identification of weaknesses that may cause vulnerabilities allows software developers, hardware engineers, and security architects to resolve them early in the development process~\cite{kimbrellkelly2025}, even before they become actual vulnerabilities and exploits.


Over time, the reliance on the NVD has revealed significant shortcomings. As vulnerability disclosures have surged both in volume and complexity, the NVD has struggled to keep up. The total CVE count rose from under 100,000 in 2015 to over 150,000 in 2020. 
and over 300,000 CVE in 2025.
Since February 2024, 
only 58\% of CVEs added to the NVD 
have CWE mappings. This pause propagated until mid-2025 and is ongoing, resulting in almost half of new CVEs each month missing enrichment metadata (CPE, CWE, CVSS). Although the NVD has announced its intention to restore its prior operational pace~\cite{nvdnews}, the delay and uncertainty have left developers and security teams in the dark for extended periods. 

\begin{figure}[!t]
    \centering\includegraphics[width=0.6\linewidth]{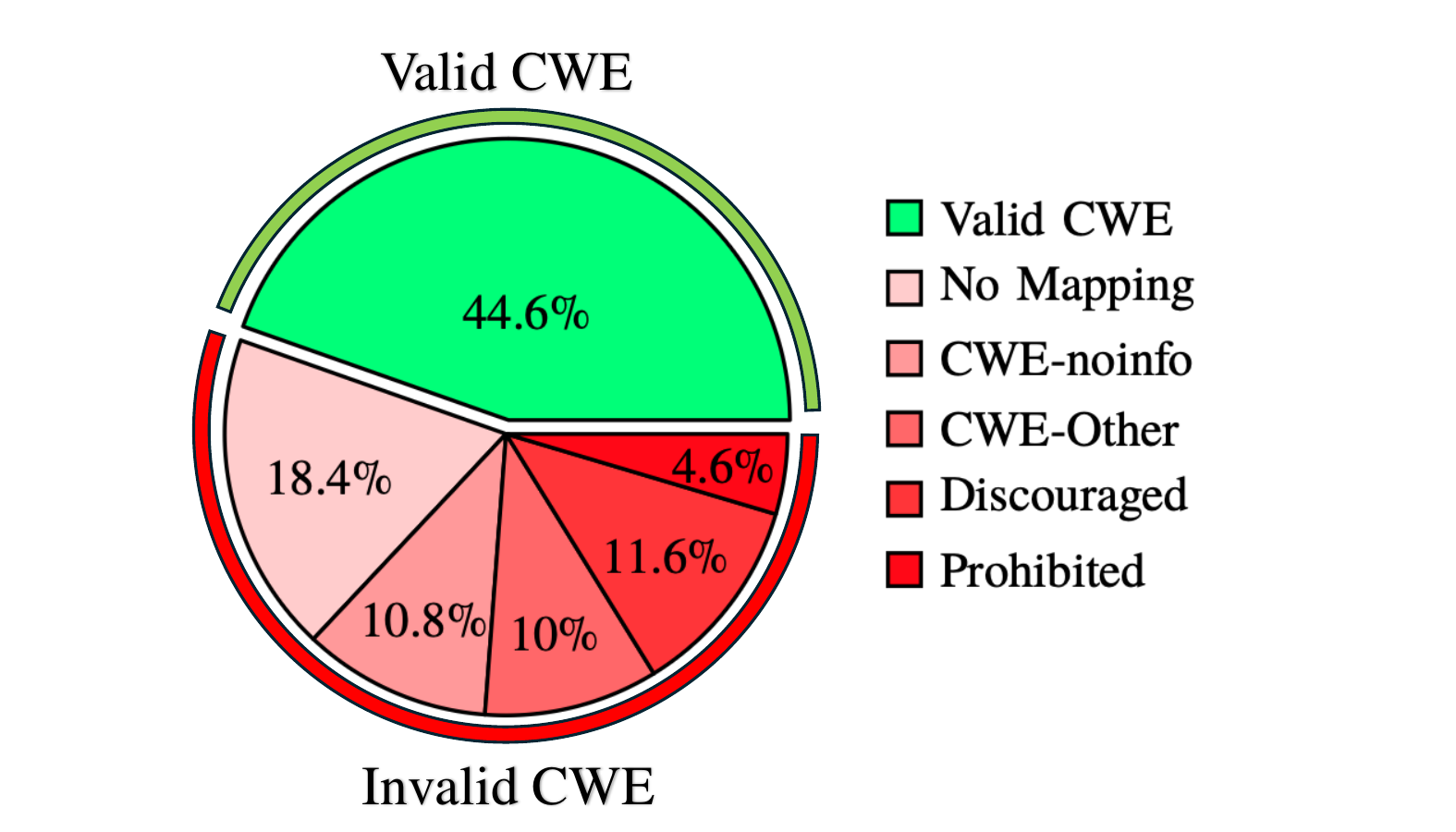}
\caption{The proportion of invalid mappings versus valid CVE-CWE mappings among 280,000+ CVEs (as of December 17, 2024). The diagram shows that valid mappings currently represent fewer than half of the entries in the National Vulnerability Database (NVD).}
\label{fig:cwe-24-dist}
\end{figure}

Despite the focus on recent disruptions caused by the enrichment pause back in February 2024, long-standing structural and operational issues with CVE metadata have persisted. Through a longitudinal analysis, revealed in this paper, we identify a largely overlooked problem within the NVD. Specifically, our evaluation of NVD entries as of December 2024 finds that 55\% of all CVEs in the NVD are mapped to \emph{invalid} CWEs. Invalid CWEs refer to entries that are \emph{Prohibited}, \emph{Discouraged}, or serve merely as placeholders (e.g., \texttt{CWE-Other}, \texttt{CWE-noinfo}), as shown in Figure ~\ref{fig:cwe-24-dist}. The Prohibited and Discouraged categories, introduced by MITRE in 2019, limit the use of overly generic or non-actionable CWE tags that lack diagnostic value. However, despite efforts to phase them out, our data shows that 5\% of CVEs submitted since 2019 still use Discouraged CWEs, indicating ongoing problems with mapping quality. 

Invalid mappings reduce the value of vulnerability databases by obscuring the true root causes of vulnerabilities, limiting the ability of security teams to detect attack patterns and hindering the design of effective mitigation strategies. Thus, poor understanding of the root cause of a vulnerability may result in the CVE being exploited in the wild. Indeed, focusing on CVEs that were added to the Known Exploited Vulnerabilities (KEV) list~\cite{cisa-kev-catalog} between  August 2021 and December 2024, we find that 99 of them were originally mapped to Discouraged or Prohibited CWE. Furthermore, 95 out of these 99 exploited vulnerabilities were remapped to the correct CWE only after the exploit was revealed. The issue is particularly critical given the widespread reliance of open-source developers, security researchers, and automated tools on the NVD as the primary source of structured vulnerability data.


\noindent \textbf{Our Contributions.}
In this work, we address the problem of automatically correcting the large number of Prohibited and Discouraged CVE-CWE mappings in the NVD, through the combination of longitudinal analysis and knowledge graph reasoning. 



First, to inform our predictions, we systematically analyze the National Vulnerability Database (NVD), showing that Prohibited mappings make up to ~5\% of all mappings in the NVD, and Discouraged CWE mappings make up to ~12\%. Through a longitudinal analysis of CVE-CWE mapping updates between 2016 and 2024, we observe that 84\% of remaps remain within the same CWE branch, often only one to two hops away within the CWE hierarchy, with a maximum of 4 hops connected by ChildOf/ParentOf relationship.
For CWEs that are Prohibited (i.e Categories or Views), we discover that the replacing CWEs are often a member of the original Category  or View. 

Building on these insights, we construct a \emph{knowledge graph ontology} that integrates CVEs, CWEs, and CPEs as interconnected entities linked by relationships (the ontology of a knowledge graph is a structured framework that defines the types of entities, their properties, and the relationships between them). Using graph embeddings trained on this structure, we introduce \fixvw, a novel method that predicts allowed CWE candidates for vulnerabilities currently mapped to Prohibited or Discouraged entries.
\fixvw\ uses only graph connections and does not require semantic information, thus providing a lightweight, scalable, and fast prediction model. We test \fixvw\ on six different candidate sets: CWE-1003 for baseline, CWE Top-25 list as considered in other studies~\cite{albanesecve2cwe,aghaei2020threatzoom}, and four candidate entity sets informed by our longitudinal study. For Discouraged mappings, we constrain predictions to children or siblings in a CWE branch; for Prohibited category/view remaps, we extract valid member CWEs and their descendants 
and fill in with nearest neighbors from the ontology when the size of the candidate entity set is too small, to reach a sufficient number of candidates.


Our model is validated under an open-world assumption (OWA) which is generally considered more realistic than the closed-world assumption (CWA) predominantly used in the evaluation of knowledge graphs~\cite{NEURIPS2022_378226e5}. Specifically, we validate \fixvw\ using real-world corrections applied on the NVD between August 2021 and December 2024, treating updated CWE remaps as ground truth. We evaluate our model's Top-10 CWE predictions on three levels of coverage metrics; exact match (correct CWE mapping from the NVD), fine-grain match (a directly related CWE in the same branch i.e. a parent or a child) and coarse grain match (any descendant from the same branch root).
We show how the fixed NVD dataset, obtained after applying \fixvw\ on the original NVD dataset, improves the accuracy of ML models relying on the NVD. In particular, replacing Prohibited/Discouraged mappings with the Top-2 predictions obtained from \fixvw\ significantly improves the performance of graph completion models. 
 We further analyze CVEs belonging to the Known Exploited Vulnerabilities Catalog~\cite{cisa-kev-catalog} and Exploit DB list~\cite{exploitdb} and show that since August 2021, 190 CVEs were mapped to Prohibited or Discouraged CWEs. Our model is able to correctly predict the root cause for 131 of them, which means that
\fixvw\ could have correctly identified the underlying weaknesses and possibly prevented these CVEs from being exploited.   

Our contributions can therefore be summarized as follows:
\begin{itemize}
    \item We conduct the first large-scale longitudinal analysis of NVD CVE–CWE mappings, revealing that more than half are invalid but systematically linked to correct CWEs through hierarchical structures. 
    \item We identify structural patterns in CWE remaps, showing that Prohibited mappings typically resolve to member CWEs, while Discouraged mappings remain within nearby branch relationships.
    \item We propose \fixvw, a lightweight and deterministic knowledge-graph embedding method that leverages CVE–CWE–CPE connectivity without requiring semantic features to correct invalid mappings.
    \item We propose and correct different sets of candidate entities for correcting the invalid CVE-CWE mappings, validating our results with correct labels taken from updated mappings in the NVD. 
        \item We validate our approach using real-world CWE remaps under an open-world assumption( OWA), and showcasing our model's performance on predicting correct mappings for CVEs with exploits.
    \item We show that the knowledge graph embedding model retrained with the top CWE labels returned by \fixvw\ significantly outperforms the original model in predicting unseen CVE-CWE triples (currently unknown, but revealed in the future), demonstrating the effectiveness of \fixvw\ in improving existing ML models based on NVD data.
\end{itemize}

The rest of this paper is organized as follows. Section \ref{sec:literature} discusses related work. Section \ref{sec:background} explains core concepts and terminology, and present the findings of our longitudinal analysis. In Section \ref{sec:fixv2w}, we introduce the \fixvw\ method, and detail its properties. We present our implementation and evaluation methodology in Section \ref{sec:methodology}.  In Section ~\ref{sec:eval}, we present our results and evaluation, followed by a discussion in Section ~\ref{sec:discussion}. This section also discusses limitations and threats to validity. We conclude the paper in Section ~\ref{sec:conclusion}. The software implementation of \fixvw\ and other artifacts can be found in our GitHub repository~\cite{nislabGithub}.

\section{Related Work}\label{sec:literature}

Knowledge graphs (KGs) are used in many domains, including medicine, history, and cybersecurity. A common issue across these domains is when the knowledge graph has missing or incorrect data. To address this issue, several studies focused on error detection and repair in knowledge graphs. For instance, Akasiadis \emph{et al.}~\cite{detectandfix2024} develop an open-source system for detecting and fixing formal inconsistencies in large real-world knowledge graphs. They integrate and extend state-of-the-art approaches for parallel KG inconsistency detection and fixing in a single framework.
Similarly, Arioua \emph{et al.}~\cite{arioua2018user} suggest a platform employing user-guided repair techniques for knowledge bases that enable user-suggested updates to resolve errors. The basic idea is that the knowledge base asks questions about possible fixes, and the user selects the correct ones until a consistent knowledge base is achieved. Although such an approach is possible, the volume of invalid data in the NVD and complexity of our knowledge graph makes it infeasible due to the required effort.

Arnaout \emph{et al.}~\cite{arnaout2022utilizing} suggest a method that repairs incorrect statements in KGs by replacing incorrect subject-predicate-object triples with likely correct ones, thus avoiding information loss. Although effective, the model relies on textual summaries that create a large overhead, unlike the lightweight knowledge graph used in \textit{FixV2W}. In~\cite{nerualkbrepair2021}, the authors propose to repair Wikidata using its edit history. The paper proposes a deep learning model that exploits edits that removed inconsistent triples in the past, to infer similar corrections in the present. This approach relies on a rich and diverse edit history, which does not exist in our case since very little corrections have been applied in the NVD, especially for Prohibited CWEs.

Analyzing the NVD has attracted considerable research attention. Some studies focus on CVSS accuracy~\cite{zhang2023flaw,kekul2022estimating} or missing metadata~\cite{hu2024cpe,kuhn2021}, while others examine the overall viability of open-source vulnerability data~\cite{dong2019inconsistencycve,jiangevaluatingos2021} and the completeness of related information. Overall, each of these studies points out issues in the NVD data such as incorrect CPE identifiers, missing or wrong CVSS vectors and scores and suggest that the NVD needs to undergo a structural change to prevent such inconsistencies. 

Highlighting the issues that affect ML model accuracy, Zhang \emph{et al.}~\cite{zhang-empirical} present a model to predict zero-day vulnerabilities using the NVD and identify limitations such as missing version information, data errors, and delays in release times. We conclude from their findings that correcting errors in the NVD is a necessary step for building predictive models. Moreover, a 2023 study by Croft \emph{et al.}~\cite{croftdataquality2023} found that between 20\% and 71\% of vulnerability labels in real-world datasets are inaccurate. Anwar \emph{et al.}~\cite{anwar2020} assess the quality of the NVD, addressing incorrect vendor information and vulnerability publication dates that affect CVSS scores. They develop an automation tool to correct these errors. While they examine missing or incorrect CVE entry information, we focus on \emph{relationships} between CVE and  CWE entries. Nguyen \emph{et al.}~\cite{nguyen2013} study vulnerabilities in Google Chrome, revealing that many originated from early versions and highlighting errors in vulnerability data. This supports our assertion that CVE-CPE mappings can be flawed, leading to incorrect enrichment metadata. Kühn \emph{et al.}~\cite{kuhn2021} apply machine learning and NLP to improve the NVD's information quality, focusing on the atomic accuracy of CVE entries (e.g., security-relevant tags and the CVSS score). In contrast, we analyze the NVD with a focus on the root causes of the vulnerabilities (namely their weaknesses), and correcting mappings between CVEs and CWES. Collectively, researchers agree that these issues cause significant impacts on downstream models, either preventing effective model training or inflating benchmark performance.  

There are also several studies automate CVE to CWE mappings using various machine learning techniques, such as natural language processing (NLP), neural networks, knowledge graphs, and large language models (LLMs). ThreatZoom~\cite{aghaei2020threatzoom} uses a hierarchical neural network and reports strong coarse-grain classification results. However, their evaluation is on unclassified CVEs using expert-annotated validation sets. \fixvw, on the other hand, focuses on correcting invalid mappings across the full NVD and achieves a 50\% exact match rate on the combined test set of Prohibited and Discouraged CWE mapping correction. CVE2CWE~\cite{albanesecve2cwe} relies on TF-IDF similarity, testing against only 50 CWEs from the 130 in CWE-1003, and lacks validation, leading to weaker accuracy compared to our graph-based method.

Shi \emph{et al.}~\cite{zpshitops} apply various KG models to the problem of CVE-CWE-CPE graph completion. We build on this insight, but further improve precision through graph-based candidate selection and by learning from the NVD’s remapping history.
VulnScopper~\cite{alfasi2024} applies the ULTRA foundation model, enhanced with OpenAI’s Ada LLM, for the same type of graph completion. Its performance depends on a multi-stage architecture with heavy reliance on LLM-generated semantics. In our paper, we show that after applying \fixvw\ on the NVD, it is possible to achieve similar accuracy without requiring complex semantic models.

CWE-GPT~\cite{madden2023vulnerability}, developed by MITRE’s CyberSecAI initiative, is a grounded LLM-based system designed to improve CWE mapping for CVEs. It combines retrieval-augmented generation with enriched CVE data and CWE ontology to reduce hallucinations and enhance accuracy. The tool supports both interactive use and bulk processing, with open-source components available. Evaluated on a large CVE dataset, CWE-GPT achieves high recall but lower precision with 95.85\% of CVEs with at least one CWE match and precision score of 0.10~\cite{cmadden-cwegpt}, reflecting its focus on generating comprehensive candidate CWE lists. 
In contrast, \fixvw\  focuses on fixing existing, invalid mappings. A key advantage of \fixvw\ is to offer deterministic and interpretable results, as it does not depend on LLMs.

Finally, Mittal \emph{et al.}~\cite{mittal2019vulnkg} develop Cyber-All-Intel, a cybersecurity knowledge graph used to link threat actors, software, and vulnerabilities. Cyber-All-Intel demonstrates that structured threat intelligence can be automatically constructed and made queryable, facilitating more effective threat hunting and situational awareness. However, while it incorporates CVE and CWE data, the system focuses on entity extraction and graph construction—not on correcting or validating vulnerability mappings.

To the best of our knowledge, our work is the first that specifically targets the correction of invalid CWE mappings in the NVD. While other efforts focus on assigning new CWEs or classifying unmapped CVEs, \fixvw\ is designed to fix existing erroneous mappings by leveraging both the structure of the CWE ontology and observed mapping evolution over time.
An earlier and shorter version of this paper appeared in~\cite{COMPSAC25}. The main novel contributions of the journal version with respect to the conference version include: a more detailed presentation of the \fixvw\ method and thorough characterization of its properties; much more extensive evaluation of the performance of  \fixvw\, including with different candidate entity sets, and with exploited CVEs; demonstration of how ML models relying on the NVD can be improved; extensive discussion of the findings, evaluation methods, and limitations of our work.

\section{Background and Longitudinal Analysis of the NVD}\label{sec:background}
This section reviews key concepts related to the NVD, CVE, CWE, and associated terminology, followed by a longitudinal analysis of NVD data that informs the design of \fixvw.

\raggedbottom
\subsection{The National Vulnerability Database and data quality}

The National Vulnerability Database (NVD) is the U.S. database of vulnerability management data and serves as the central source of vulnerability information for many other databases and tools. This database enables automation of vulnerability management, security measurement, and compliance. The NVD includes CWE identifiers representing software security flaws, product names in CPE format, and impact metrics such as CVSS scores in its CVE metadata. The NVD has been operating on the enrichment of publicly disclosed vulnerabilities with the help of Certified Numbering Authorities (CNAs). CNAs are verified submitters of vulnerabilities that report on behalf of themselves and others. The NVD relies on Certified Numbering Authorities (CNAs), organizations such as Apache, Red Hat, and Microsoft, which submit and enrich vulnerability reports with CWEs, CPEs, and CVSS scores.

Given the volume of vulnerabilities submitted, the NVD team struggles to keep up with vulnerability enrichment, which negatively affects security analysis efforts. It was reported in~\cite{paloaltoKEV} that many CVEs are already exploited by the time they make it into the database, with 14\% of the exploits being zero-day vulnerabilities and 80\% of public exploits being published before the CVEs are. These challenges underscore the urgent need for automated vulnerability analysis to speed up the process and increase the accuracy of vulnerability records. Current efforts to automate this process struggle from low quality of vulnerability data, therefore it is critical to fix incorrect mappings in these databases.

\subsection{CVE, CWE and mapping metadata}

The Common Vulnerabilities and Exposures (CVE)~\cite{cve} is a system developed by MITRE to catalog publicly known vulnerabilities. Each newly reported vulnerability receives a unique CVE ID. Submissions are typically made by a Certified Numbering Authority (CNA), who is often the vendor responsible for the affected product or a researcher/individual submitting vulnerabilities through established reporting processes.

The Common Weakness Enumeration (CWE)~\cite{cwe}, also managed by MITRE, is a classification system for software and hardware weaknesses that can lead to vulnerabilities under certain conditions. Each CWE entry is assigned a CWE ID, which may refer to a specific weakness, a broader category, or a view.

\paragraph{Category and View} Categories are groups of related weaknesses that share a common trait. A category is not a weakness, but rather a structural item that helps users find weaknesses that share the stated common characteristic. Currently, the CWE database lists 374 categories, though some are empty or obsolete. 
Views are a subset of CWE entries that provide a way of examining CWE content. Views organize weaknesses based on different criteria, such as inclusion in an annual CWE Top-25 list curated by MITRE, which lists the most common and dangerous CWE in the past year(s)~\cite{cwe-25}. Currently, 51~views exist, each serving different purposes. In our work, we focus on the \emph{CWE-1003 view: Weaknesses for Simplified Mapping of Published Vulnerabilities} since it is being used by the NVD for root cause mapping of vulnerabilities. This view was first curated in 2016 and ever since, NVD exclusively maps CVEs to CWEs belonging to this view. An example category, \emph{CWE-189: Numeric Errors}, is shown in Table ~\ref{tab:cwe-cate} along with the members of this category.

\begin{table}[t]
    \centering
    \caption{CWE-189 ``Numeric Errors'' category, and its members that are allowed CWE. The CWE in bold are included in the CWE-1003 view, and are used for root cause mapping of CVEs by the NVD.}
    \label{tab:cwe-cate}
    \begin{tabular}{c|c}
         Allowed CWE & CWE-189 ``Numeric Errors'' Category Members \\
         \hline
        CWE-128 & Wrap-around Error \\
        \textbf{CWE-190} & \textbf{Integer Overflow or Wraparound} \\
        \textbf{CWE-191} & \textbf{Integer Underflow (Wrap or Wraparound)} \\
        \textbf{CWE-193} & \textbf{Off-by-one Error} \\
        \textbf{CWE-369} & \textbf{Divide By Zero} \\
        \textbf{CWE-681} & \textbf{Incorrect Conversion between Numeric Types} \\
        CWE-839 & Numeric Range Comparison Without Minimum Check \\
        CWE-1135 & Incorrect Bit-wise Shift of Integer \\
        CWE-1139 & Insufficient Precision or Accuracy of a Real Number \\
        CWE-1389 & Incorrect Parsing of Numbers with Different Radices\\
        \hline
    \end{tabular}
\end{table}

\paragraph{Prohibited and Discouraged Tags}
Prior to 2019, any type of CWE could be used to describe the root cause of a vulnerability. Since then, MITRE has introduced the Prohibited and Discouraged tags:
\begin{itemize}
    \item \textbf{Prohibited CWEs} (such as categories and views, and deprecated entries) are no longer allowed for root cause mapping because they do not represent valid weaknesses; instead, they serve as collections of related CWEs. Table ~\ref{tab:cwe-cate} gives an example of a category and its associated members.

    \item \textbf{Discouraged CWEs} are weakness entries that are often too broad to clearly explain the root cause of a vulnerability. CNAs are urged to use more specific CWE entries whenever possible to more accurately reflect how a vulnerability was introduced.
\end{itemize}

These two classes of invalid CWE mappings affect the mapping data and create confusion in the ontology, therefore we aim in this work to correct those CVE that are mapped to Prohibited or Discouraged CWE, whenever possible. We use the results from our longitudinal analysis to inform the predictions, rather than blindly querying for the entire CWE database, or the CWE-1003 view. We present comparisons of the prediction performance with each of these candidate entity sets and rationale for using them in Section ~\ref{sec:candidate}.

\paragraph{The CWE-1003 view} NVD uses the CWE-1003 view for providing root cause information to CVEs. The CWE-1003 view has stayed the same while the number of CWE expanded, therefore this list has obvious shortcomings. The CWE-1003 view, finalized in 2019, contains 130 CWEs but hasn’t kept pace with the expanding CWE database (940+ entries), leading to frequent use of placeholder CWEs like \texttt{CWE-Other} and \texttt{CWE-noinfo} in over 25\% of CVEs as of 2025. 

\begin{figure}[t]
    \centering
    \includegraphics[width=.75\linewidth]{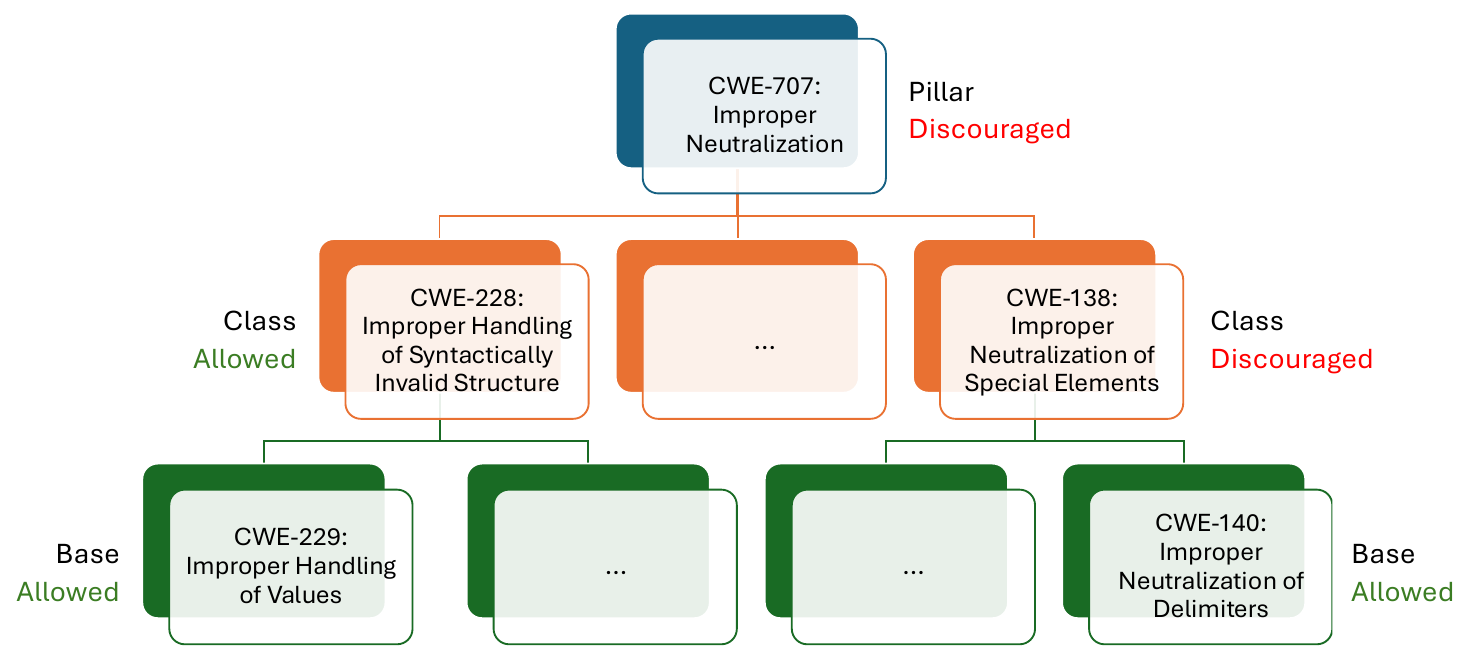}
    \caption{A slice of the CWE Hierarchy, showing parent-child relationships.
    CWE-707 and CWE-138 are both labeled \textit{Discouraged} by MITRE for CVE root cause mapping, due to lack of detail. Weaknesses are at deeper levels of the hierarchy are more detailed.}
    \label{fig:hierarchy}
\end{figure}

The CWE structure is hierarchical, where entries may be related with Parent/Child relationships, which shows weaknesses at different levels of abstraction. Typically, a CWE hierarchy starts with a Pillar, followed by a less general Class or Base weakness, and ends with the most detailed Variant weakness~\cite{cwehierarchy}. All Pillar type CWEs and some of the Class type CWEs are Discouraged for CVE mapping. Some hierarchies do not include variants, but almost all include Class and Base. Figure~\ref{fig:hierarchy} illustrates an example of this hierarchy. 

\subsection{Longitudinal analysis of the NVD}\label{sec:Longitudinal}

Longitudinal analysis is a tool that aids in understanding how patterns evolve over time, revealing trends that persist or changes that are introduced~\cite{donald2006longitudinal}. We conduct a longitudinal analysis to understand how the mapping behavior of CVE root causes changed over time and what trends we can learn from the NVD. In the following, we present our results and key takeaways obtained by the longitudinal analysis including providing insight on how to curate the candidate sets for informed prediction of correct CVE-CWE mappings.

\paragraph{Discouraged and Prohibited CWE Mappings} 

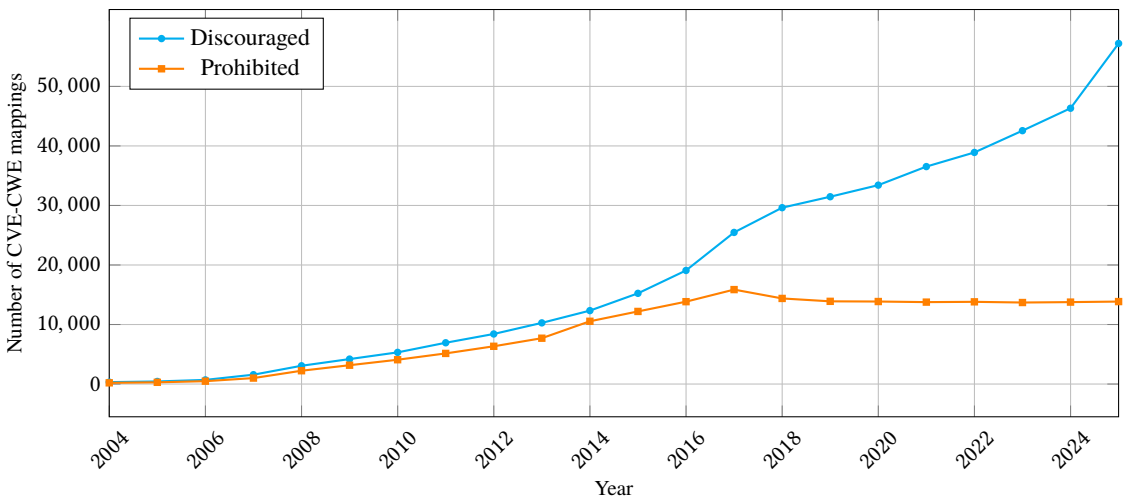
\begin{figure*} [t]
\begin{tikzpicture} 
\begin{axis}[
    width=15cm,
    height=7cm,
    xlabel={Year},
    xlabel style={yshift=-6pt},
    ylabel={Number of CVE-CWE mappings},
    xmin=2004, xmax=2025,
    ymajorgrids=true,
    xmajorgrids = true,
    grid style=solid,
    legend style={at={(0.02,0.98)},anchor=north west,legend columns=1},
    label style={font=\small},
    scaled y ticks = false,
    ytick={0, 10000,20000,30000,40000,50000}, 
    y tick label style={/pgf/number format/.cd, fixed, 1000 sep={,}}, 
    x tick label style={rotate=45, /pgf/number format/.cd, 1000 sep={}}, 
]

\addplot[
    color=cyan,
    thick,
    mark=*,
    mark options={scale=0.5,fill=cyan}
] coordinates {
(2025,57214)
(2024,46332)
(2023,42565)
(2022,38906)
(2021,36529)
(2020,33412)
(2019,31466)
(2018,29626)
(2017,25470)
(2016,19082)
(2015,15238)
(2014,12334)
(2013,10276)
(2012,8411)
(2011,6935)
(2010,5321)
(2009,4194)
(2008,3069)
(2007,1571)
(2006,696)
(2005,434)
(2004,318)
};
\addlegendentry{Discouraged}

\addplot[
    color=orange,
    thick,
    mark=square*,
    mark options={scale=0.5,fill=orange},
] coordinates {
(2025,13854)
(2024,13764)
(2023,13695)
(2022,13811)
(2021,13762)
(2020,13862)
(2019,13894)
(2018,14381)
(2017,15866)
(2016,13837)
(2015,12197)
(2014,10545)
(2013,7700)
(2012,6343)
(2011,5145)
(2010,4084)
(2009,3164)
(2008,2236)
(2007,1001)
(2006,481)
(2005,290)
(2004,202)
};
\addlegendentry{Prohibited}

\end{axis}
\end{tikzpicture}
\caption{Cumulative counts of Prohibited and Discouraged mappings over  time. Although Prohibited Mappings slightly decrease over time, Discouraged CWE-CVE mappings are increasing. }
    \label{fig:cve-unmapped}
\end{figure*}

We record the total number of Discouraged and Prohibited mappings over time, recovering the state of the NVD at each year using the NVD CVE Change History API. Figure \ref{fig:cve-unmapped} reveals the cumulative count of Discouraged and Prohibited mappings at the end of each year since 2004. Our analysis reveals a steady number of Prohibited mappings (about 16,000 since 2018) and a growing number of Discouraged mappings (exceeding 55,000 in 2025). 

\pgfplotstableread[row sep=\\,col sep=&]{
    year & Total Unmapped & CWE-Other & CWE-noinfo & Empty \\
    2008 & 13 & 2 & 8 & 2.4 \\
    2009 & 17.1 & 4.3 & 10.3 & 2 \\
    2010 & 23 & 7.1 & 12.9 & 2.9 \\
    2011 & 21 & 4 & 12.1  & 4.9 \\
    2012 & 26 & 3.9 & 14.9 & 7.1 \\
    2013 & 25.1 & 2.4 & 14.7 & 8 \\
    2014 & 18.6 & 3.8 & 8.3 & 6.4 \\
    2015 & 22.1 & 6.3 & 8.1 & 7.7 \\
    2016 & 22.8 & 2.7 & 7.9 & 12.2 \\
    2017 & 24.6 & 0 & 11.5 & 13 \\
    2018 & 18.3 & 0.1 & 11.8 & 6.5 \\
    2019 & 23.6 & 1.2 & 14.9 & 7.4 \\
    2020 & 26.2 & 2.1 & 17.5 & 6.6 \\
    2021 & 19.2 & 3.9 & 13.3 & 2 \\ 
    2022 & 16.2 & 1.9 & 12.4 & 1.9 \\
    2023 & 48 & 2.2 & 13 & 34 \\ 
    2024 & 52 & 1.3 & 9.3 & 51\\
    }\FigurefourData
\pgfplotstableread[row sep=\\,col sep=&]{
    year & Total Invalid & Empty & CWE-Other & CWE-noinfo & Discouraged & Prohibited \\
    2008 & 97.32620320855615 & 91.44385026737967 & 0.0 & 0.5347593582887701 & 3.7433155080213902 & 1.6042780748663101 \\
    2009 & 83.33333333333334 & 76.11111111111111 & 0.0 & 0.5555555555555556 & 5.555555555555555 & 1.1111111111111112 \\
    2010 & 77.77777777777779 & 64.93055555555556 & 0.0 & 0.3472222222222222 & 12.5 & 0.0 \\
    2011 & 72.51585623678646 & 55.81395348837209 & 0.6342494714587738 & 2.1141649048625792 & 13.10782241014799 & 0.8456659619450317 \\
    2012 & 79.83870967741937 & 63.03763440860215 & 0.6720430107526881 & 2.6881720430107525 & 13.306451612903226 & 0.13440860215053763 \\
    2013 & 72.38999137187231 & 54.5297670405522 & 0.6902502157031924 & 0.7765314926660914 & 15.099223468507335 & 1.2942191544434858 \\
    2014 & 70.59748427672956 & 50.5503144654088 & 0.31446540880503143 & 0.9433962264150944 & 14.937106918238994 & 3.852201257861635 \\
    2015 & 67.37633061991234 & 49.53036944270507 & 0.06261740763932373 & 0.6887914840325611 & 14.08891671884784 & 3.005635566687539 \\
    2016 & 79.69502407704654 & 53.93258426966292 & 0.20064205457463885 & 0.8828250401284109 & 14.526484751203853 & 10.152487961476725 \\
    2017 & 66.10871991072852 & 39.45480631276901 & 0.11158935118762953 & 5.037462139327276 & 20.293320580264627 & 1.2115415271799777 \\
    2018 & 41.214165261382796 & 9.904440697020798 & 0.1461495222034851 & 11.202922990444069 & 19.763912310286678 & 0.1967397414277684 \\
    2019 & 36.2511353315168 & 9.54246139872843 & 1.1977747502270664 & 14.220027247956404 & 10.757266121707538 & 0.5336058128973661 \\
    2020 & 39.61252703846516 & 11.760556757265118 & 1.9984952506348161 & 16.42527978933509 & 9.428195241230133 & 0.0 \\
    2021 & 42.50107066381156 & 16.548179871520343 & 3.6616702355460387 & 13.143468950749465 & 9.143468950749465 & 0.004282655246252677 \\
    2022 & 44.0983180484907 & 22.214383841384176 & 4.01737645267887 & 12.174655626926075 & 5.680763375784354 & 0.011138751717224223 \\
    2023 & 52.02231054407 & 30.79389465949701 & 2.287832737717511 & 13.833873284125447 & 5.103369960923149 & 0.0033399018068868773 \\
    2024 & 56.129430526261636 & 42.65137095529605 & 1.5535778589433613 & 9.88991203251196 & 2.0139924893255827 & 0.020577190184680284 \\
    }\FigureDataUnmapped

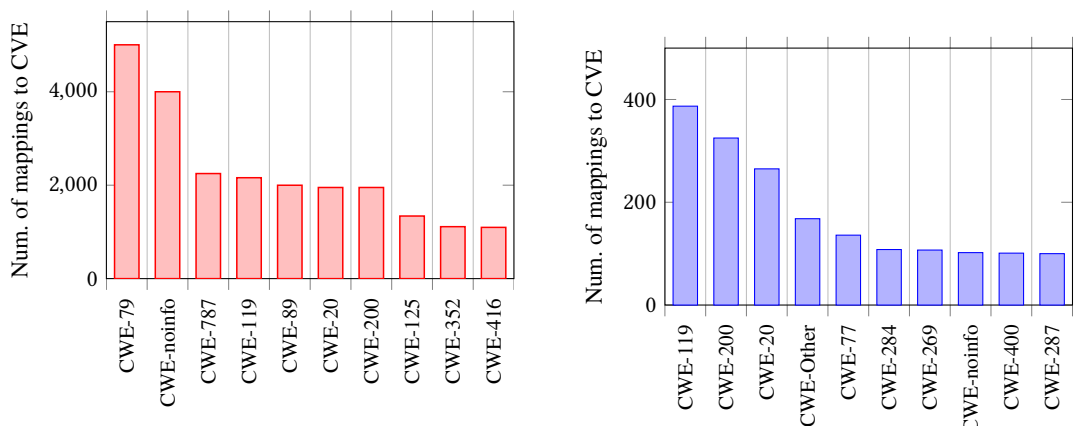
\begin{figure}[tp]
\begin{subfigure}{.45\textwidth}
    \begin{tikzpicture}
    \pgfplotsset{width=7cm,height = 5cm, compat=1.9}
    \pgfplotsset{xticklabel={\tick},scaled x ticks=false}
    \pgfplotsset{plot coordinates/math parser=false}
    \begin{axis}[
        ybar,
        ymin=0, ymax=5500,
        xticklabel style={rotate=90},
        xticklabel style={font=\small},
        symbolic x coords={CWE-79, CWE-noinfo, CWE-787, CWE-119,  CWE-89, CWE-20, CWE-200, CWE-125, CWE-352, CWE-416, CWE-Other},
    	ylabel=Num. of mappings to CVE,
    	ybar interval=0.6,
        enlargelimits=false
    ]
    \addplot [draw = red,
        line width = .2mm,
        fill = pink]
    	coordinates{ (CWE-79,5005) (CWE-noinfo,4000) (CWE-787,2249)(CWE-119, 2160)(CWE-89, 1999) (CWE-20,1951)(CWE-200, 1951)(CWE-125,1340)(CWE-352,1112)(CWE-416,1097)(CWE-Other, 1094) };
    \end{axis}
    \end{tikzpicture}
    \caption{Most occurring newly mapped CWEs between 2016-2024. From left to right, the Top-3 CWE are CWE-79: Improper Neutralization of Input During Web Page Generation ('Cross-site Scripting'), CWE-787: Out-of-bounds Write and CWE 119: Improper Restriction of Operations within the Bounds of a Memory Buffer.}
    \label{fig:CWE-long-new}
\end{subfigure}%
\hspace{.05\textwidth}
\begin{subfigure}{.45\textwidth}
    \begin{tikzpicture}
    \pgfplotsset{width=7cm, height = 5cm, compat=1.9}
    \pgfplotsset{xticklabel={\tick},scaled x ticks=false}
    \pgfplotsset{plot coordinates/math parser=false}
    \begin{axis}[
        ybar,
        ymin=0, ymax=500,
        xticklabel style={rotate=90},
        xticklabel style={font=\small},
        symbolic x coords={CWE-119,  CWE-200,  CWE-20,  CWE-Other, CWE-77, CWE-284, CWE-269, CWE-noinfo, CWE-400,CWE-287, CWE-863},
    	ylabel=Num. of mappings to CVE,
    	ybar interval=0.6,
        enlargelimits=false
    ]
    \addplot 
    	coordinates{ (CWE-119, 387) (CWE-200, 325) (CWE-20, 265) (CWE-Other, 168) (CWE-77, 136) (CWE-284, 108)(CWE-269, 107)(CWE-noinfo, 102)(CWE-400, 101)(CWE-287, 100)(CWE-863, 100)};
    \end{axis}
    \end{tikzpicture}
    \caption{Most removed CWEs between 2016-2024, from left to right, Top-3 are CWE 119: Improper Restriction of Operations within the Bounds of a Memory Buffer, CWE-200: Exposure of Sensitive Information to an Unauthorized Actor, and CWE 20: Improper Input Validation.}
    \label{fig:CWE-long-old}
\end{subfigure}
\caption{Longitudinal analysis between 2016-2024 reveals the most remapped CWE, showing (left) certain CWE were added that include more detail and (right) certain more general CWE were removed from mappings more often.}
\end{figure}

\paragraph{The use of CWE-Other and CWE-noinfo}
A total of 4000 updates between July 2016 and December 2024 were remaps into \texttt{CWE-noinfo}. In fact, Figure~\ref{fig:CWE-long-new} shows that among the Top-three CWEs used to provide root cause mappings to CVEs, \texttt{CWE-noinfo} ranks second. This clearly shows that the CWE-1003 view falls short in covering all types of CWEs, and officials must update the list to cover more representative CWEs. Ever since the CWE-1003 was last updated in 2019, more than 70 new CWEs have been added to the CWE database. 

\paragraph{Commonly misused CWEs}
Figure ~\ref{fig:CWE-long-old} shows that certain CWEs are more often misused, in particular CWE-119, CWE-200, CWE-20 and so on. All of these CWEs are now labeled as Discouraged for vulnerability mapping since they lack specificity, as do not say much about the root cause of a CVE. Although some CWEs are misused often, they cannot be excluded from the candidate sets because they appear often in the newly mapped CVEs, as shown in Figure~\ref{fig:CWE-long-new}. Furthermore, NVD allows use of these CWE in root cause mapping, when a more detailed alternative is not available, thus they cannot be avoided completely. We present and evaluate our approach for fixing invalid CWEs in Section~\ref{sec:fixv2w}.

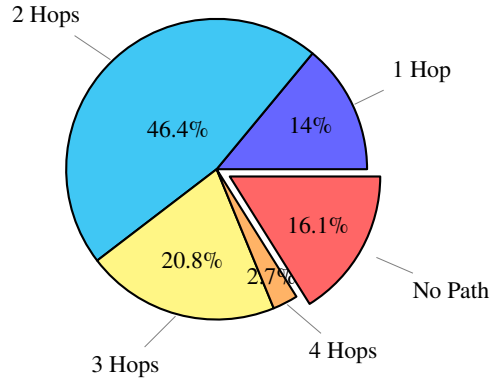
\begin{figure}
    \centering
    \begin{tikzpicture}
        \pie[radius=2, text = pin, explode={0,0,0,0,0.2}]{14/1 Hop, 46.4/2 Hops, 20.8/3 Hops, 2.7/4 Hops, 16.1/No Path}
        \end{tikzpicture}
        \caption{Distance between old CWEs and new CWEs in CVE-CWE mapping updates, between 2016-2024. Most remaps end up in CWEs that are just a few hops away, informing the design of \fixvw. }
        \label{fig:cwe-remaps}
\end{figure}

\paragraph{Updated CWE mappings that are related}
Among the CVE remaps to valid CWEs, 60\% result in a new CWE that is at most two hops away from the original, and over 80\% remain within the same branch of the CWE hierarchy. Figure~\ref{fig:cwe-remaps} shows the distance between old CWE and new CWE pairs remapped between 2016 and 2024. Most corrections are minor (one or two hops in the hierarchy), meaning a direct child is the most likely candidate for the correction of a Discouraged CWE. Distance is measured in any direction, and includes going up to the starting CWE's parent and then going down to another descendant. It is important to take into consideration such cases when searching for good candidates for CWE correction task in \fixvw. Exceptions exist and complicate full automation, and in those cases expert analysis plays a crucial role. 

This suggests that the correct CWE is often reachable from the original incorrect mapping by traversing related weakness chains, typically moving from a broader (parent) CWE to a more specific descendant CWE within the same branch. Table~\ref{tab:CWE_pairs} shows the most observed remap pairs. Strikingly, 38\% of remaps are just removals of old CWEs and replacing them with \texttt{CWE-noinfo}.
Yet for the remaining 62\% of remaps, most stay within the same CWE chain or are a member of the old CWE category (if the said CWE category has any members at all.) 

\paragraph{Summary} Our longitudinal analysis of the NVD indicates that while CNAs have occasionally corrected CWE mappings over time, such updates remain relatively rare. We identified several trends in these remaps, which inform our approach to select candidate entities for each type of invalid CWE mapping. For example, when a CVE is originally mapped to \emph{CWE-189: Numeric Errors}, it is often more effective to search within the members of that category rather than across the entire CWE-1003 view. However, this trend is not universal. In some cases, the initial mapping is substantially unrelated to the true root cause, complicating correction. Some invalid mappings are significantly misaligned with the actual root cause. For instance, CVE-2021-0572 was initially assigned \emph{CWE-287: Improper Authentication}, whereas the correct weakness, later updated in the NVD, is \emph{CWE-732: Incorrect Permission Assignment for Critical Resource}. 
Guided by these observations, we developed the \fixvw\ algorithm, presented in the next section, to search for corrected CWE mappings within the same branch or among members of the same category and view.

\begin{table}
\caption{The most observed Old CWE to New CWE pairs for the same CVE Record, that were updated between 2016 and 2024. Pairs are either in the same chain for valid CWE, or the new CWE is a member of the old CWE category, if the category has any members.}
    \label{tab:CWE_pairs}
    \centering
    \begin{tabular}{|m{4cm}|m{2cm}|m{2cm}|m{2cm}|}
    \hline
        \textbf{CWE ID Pair} & \textbf{Occurrences} & \textbf{Same Branch} & \textbf{Is Member} \\
        \hline
        CWE-Any - CWE-noinfo  & 38\% & - & - \\
        \hline
        CWE-119 - CWE-787  & 6\% & Yes & - \\
        \hline
        CWE-77 - CWE-78  & 3\% & Yes & - \\
        \hline
        CWE-119 - CWE-125  & 2\% & Yes & - \\
        \hline
        CWE-264 - CWE-732 & ~1\% & - & No$^*$ \\
        \hline
        CWE-255 - CWE-522 & ~1\% & - & Yes \\
        \hline
        CWE-264 - CWE-269 & ~1\% & - & No$^*$ \\
        \hline
        CWE-416 - CWE-787 & ~1\% & Yes & - \\
        \hline
        CWE-190 - CWE-787 & ~1\% & Yes & - \\
        \hline
        CWE-284 - CWE-732  & ~1\% & Yes & - \\
        \hline
        CWE-399 - CWE-772  & ~1\% & - & Yes \\
        \hline
    \end{tabular}
    \smallskip
    
    \footnotesize
    $^*$ CWE-264: Permissions, Privileges, and Access Controls is a category with no members.
    \end{table}


\section{FixV2W: Automated Prediction and Correction of CVE-CWE mappings}\label{sec:fixv2w}

In this section, we present \fixvw, a lightweight and structured approach to correct CVE–CWE mappings. \fixvw\ is a prediction model that uses a knowledge graph (KG) embedding approach to infer correct mappings for vulnerabilities previously linked to invalid CWE entries. We leverage existing historical mapping data along with informed predictions based on CWE hierarchy and known relationships. In our longitudinal analysis of the National Vulnerability Database (NVD), we indeed observed a consistent pattern: invalid mappings, specifically those labeled as Prohibited or Discouraged, are often replaced with more specific descendant CWEs (i.e., Members or Children), or with related weaknesses within the same branch of the CWE hierarchy. We incorporate this observation into our method by intentionally selecting candidate CWE sets that reflect these likely replacements. 


\subsection{Knowledge Graph Embedding}\label{sec:kg-embed}

\begin{figure}[]
    \centering
    \includegraphics[width=\textwidth]{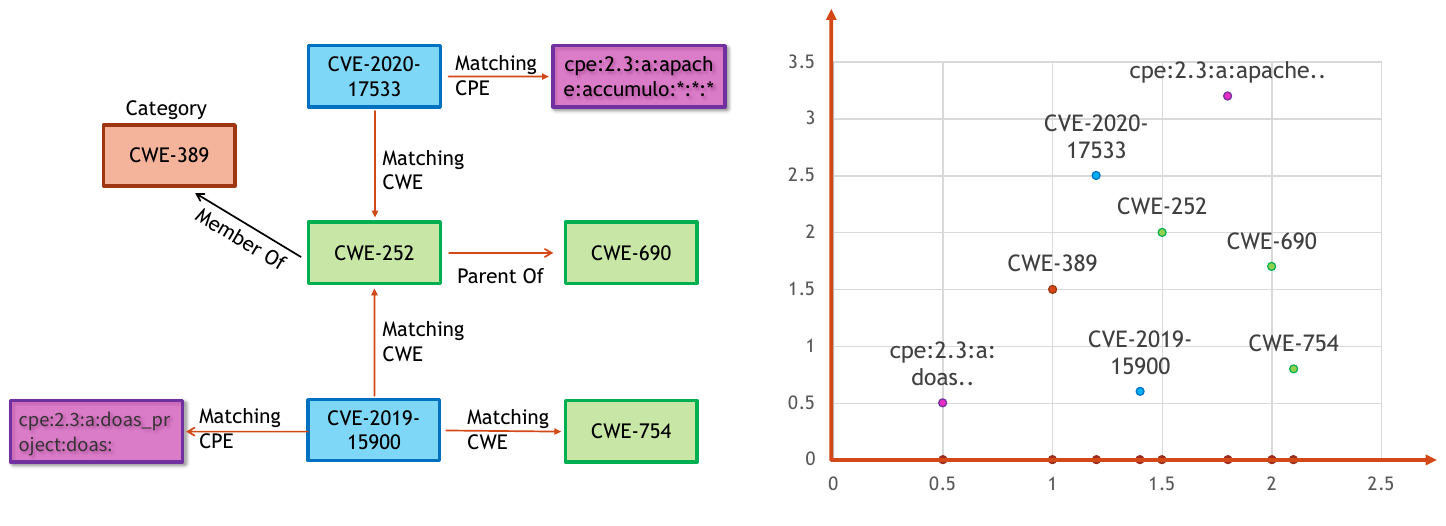}
    \caption{Knowledge Graph ontology example (left) and translation of this ontology to a 2D embedded vector space (right). Nodes that are close in the vector space are more strongly related than those that are far. Color code: CVE nodes (blue), CWE nodes (green), CPE nodes (purple), CWE category and view nodes (orange).}
    \label{fig:kg-vector}
\end{figure}

We begin by outlining the terminology and foundational concepts related to knowledge graph construction and embedding.
A Knowledge Graph (KG) is a structured representation of real-world entities and their relationships, where entities are represented as nodes and relationships as edges in a graph. This structure allows for the semantic representation of domain knowledge and supports inference, querying, and reasoning tasks. In our work, entities include CVEs (vulnerabilities), CWEs (weaknesses), and CPEs (systems, software, and packages). Relationships include \texttt{MatchingCPE}, \texttt{MatchingCWE}, and relationships between CWEs like \texttt{HasMember} and \texttt{ParentOf}.  Figure~\ref{fig:kg-vector} illustrates a slice of the knowledge graph.

The fundamental unit in a KG is a triple of the form (subject, predicate, object), such as 
(\texttt{CVE-2020-17533}, \texttt{MatchingCWE}, \texttt{CWE-252}). These triples are used as both the training input for our embedding model and the basis for downstream predictions. Each triple represents a fact in the KG. The \emph{embedding model} of a knowledge graph transform entities and relations into continuous vector spaces while preserving the graph’s structure and semantics. This enables predictive tasks such as link prediction. In our case, the embedding model is used to predict a more accurate CWE for a given CVE when the existing mapping is invalid or suboptimal.

\subsection{Formal Description}

For the purpose of our analysis, the list of vulnerabilities in the CVE database training dataset is denoted as $V=\{v_1, v_2\ldots,v_n\}$. The CWE list is denoted as  $W =\{w_1, w_2,\ldots,w_k\}$ and represents the set of all existing CWEs. The NVD's CWE slice (i.e., the CWE-1003 view) is defined as $W^{NVD}$. Clearly, $W^{NVD} \subset W$. We define weaknesses in the invalid CWE list as $W^P$ (note that $W^P$ can be defined for Discouraged, Prohibited and other reference CWEs separately). The members of this list are $w \in W^P$.  We denote by $V^P =\{ v \in V \}$ the set of CVEs $v$ that are mapped to a Prohibited CWE.
For each $v \in V^P$, we determine a set of CWEs, denoted $W^A(v)$, to which $v$ is allowed to be remapped.

The $\texttt{Predict}(c)$ function in Algorithm~\ref{alg:cap} is a 
knowledge graph operation that 
queries the embedding model with a candidate triple $c=(v,\texttt{MatchingCWE},w)$, where $w \in W^A(v)$, and returns the score computed for this triple.
Then the candidates are ordered based on the scores.

It is important to select the candidate list of CWEs $W^A(v)$ carefully. Indeed a ``popular'' CWE, one that many CVEs are mapped to, may end up at the Top-rank, due to its centrality. In the next subsection, we discuss the tradeoffs involved in the selection of candidates and discuss different the pros and cons of different approaches. Through this approach, we demonstrate that strategically reducing the candidate CWE space, guided by the structure of the CWE hierarchy and observed history of corrections leads to significantly better accuracy in correcting invalid CVE–CWE mappings compared to baseline candidate sets (e.g., CWE-1003 and CWE Top-25).


The \fixvw\ algorithm returns the reordered list of CWE candidates $W^A(v)$ in descending order of likelihood (based on the scores), with the Top-rank indicating the most probable match. 
The pseudo-code of this process is explained in Algorithm ~\ref{alg:cap}, which outlines the process for ranking the set of replacement candidates for each invalid CVE-CWE mapping.

\begin{algorithm}[!tp]
\caption{FixV2W}\label{alg:cap}
\begin{algorithmic}[1]
\State Determine $V^P =\{ v \in V \}$ such that CVE $v$ is mapped to a Prohibited CWE. 
\Procedure{FixV2W}{$V^P$}
\ForAll{$v \in V^P$}:
    \State Determine set of allowed CWEs $W^A(v)$ for CVE $v$
    \ForAll{$w \in W^A(v)$}:
        \State Form  candidate triple $c =(v,\text{MatchingCWE}, w)$
        \State Calculate score $s$ = $\texttt{Predict}(c)$
    \EndFor   
    \State Reorder elements in $W^A(v)$ in descending order of the score~$s$  found for each $w \in W^A(v)$
\EndFor  \\
\Return{$W^A(v)$ for each $v \in V^P$}
\EndProcedure
\end{algorithmic}
\end{algorithm}

\paragraph{Complexity Analysis of FixV2W}

To determine the set of Prohibited (or Discouraged) mappings in line 1, we iterate through mappings in the NVD and determine $V^P =\{ v \in V \}$, this is linear complexity and can be expressed as $O(|V|).$ The main loop in lines 3-10  runs for each $v \in V^P$. 
The inner loop runs for each $w \in W^A(v)$, where $|W^A(v)| = O(|W|)$ for each $v \in V^P$. 
To determine the candidate set $W^A={w \in V}$, we traverse the CWE hierarchy subgraph, which takes $O(|W|)$ in the worst case.
Forming a triple $c=(v,w)$ takes constant time for given $v$ and $w$. 
We cannot easily express the complexity  of the \texttt{Predict(c)} function, since it depends on parameters that are specific to the KG and its embedding model. We will therefore assume that each query takes $O(T)$ time, where $T$ is a function of the KG parameters. 
In practice, using PyTorch, the runtime of a query is on the order of one millisecond, as shown in Section~\ref{sec:methodology}.
Therefore, the total complexity for steps 3-10 is $O(|V|\cdot|W|\cdot T)$. The reordering of the elements of $W^A(v)$ in line 11 takes $O(|W|\cdot \log(|W|)$ with high probability, using an efficient algorithm like Quicksort. The total complexity is therefore
$$
O(|V|) + O(|V^P|\cdot|W|\cdot T) + O(|W|\cdot \log(|W|) = O(|V|+|V^P|\cdot|W|\cdot T + |W|\cdot \log(|W|)).
$$
This complexity is practical, considering the number of CVEs, CWEs, and Prohibited/Discouraged CVE-CWE in the NVD, and the short time needed to run a query.

\subsection{Candidate Selection Trials}\label{sec:candidate}
Effective candidate selection is crucial to achieve a balance between prediction precision and recall. Providing too many candidates dilutes the model focus leading to irrelevant Top-suggestions, while too few candidates risk excluding correct mappings. We next introduce several approaches, which we evaluate in Section~\ref{sec:eval}. A summary of the advantages and limitations of these approaches is provided in Tables~\ref{tab:candidates_base}, \ref{tab:candidates_proh} and \ref{tab:candidates_disc}.



\subsubsection{Baseline candidate entities}
\begin{table}[h]
\centering
\caption{\fixvw\ Baseline Candidate Types with Advantages and Limitations}
\label{tab:candidates_base}
\begin{tabular}{p{3.5cm}p{5.5cm}p{5.5cm}}
\toprule
\textbf{Candidate Set Type} & \textbf{Advantages} & \textbf{Limitations} \\
\midrule
\textbf{CWE-1003 Baseline} & Ensures all possible CWE are evaluated for, increases coverage. & Too many candidates introduce some issues such as unrelated but popular CWE ranking high due to node centrality.  \\
\hline
\textbf{CWE Top-25 Overlay} & Captures high-impact, frequently remapped CWEs that are most often used in root cause mapping of CVEs.  & Adds candidates that are not conceptually relevant, excludes many CWE that are relevant. Also highly affected by node centrality. \\
\hline
\end{tabular}
\end{table}%

These two candidate entity sets, CWE-1003 view and CWE Top-25 view, are the most general approach and serve the purpose of providing a baseline for the improvements that \fixvw\ introduces. CWE-1003 view, encompassing all possible CWE for mapping of vulnerabilities in the NVD, increases coverage, while certainly decreasing the concentration of the predictions. We see that there are several CWE with many connections, that appear in the returned result regardless of their relevance, due to their node centrality being very high. Node centrality in a knowledge graph refers to the most 'important' or influential nodes within the KG, which are determined by having many connections, being on many shortest paths and having connections to other influential entities~\cite{kg_central}. Similarly, Top-25 view curated each year from the most occurring weakness of the past few years, includes the most influential nodes with high node centrality, while excluding relevant yet less occurring weaknesses from the candidate set.


Some examples of central CWE nodes are \emph{CWE-79: Cross-site Scripting} and \emph{CWE-20: Improper Input Validation}, which almost always appear at Top-ranks for each CVE that we query for, when using the baseline candidate entity sets. This centrality, while sometimes useful, in our case has a negative effect in the accuracy of the predictions.

\subsubsection{CWE Branch Hierarchy}


Following the branch hierarchies and our findings from longitudinal studies, for Discouraged mappings, we use their children in the CWE-1003 view to select candidates. This allows us to query for more detailed mappings instead of the generic CWE, aiming to improve precision in CVE-CWE mappings, preserving the relevance of the original mapping. One setback of this method is that there are less than sufficient number of candidates (i.e. the old CWE mapping only has a few children in the CWE-1003 view) for a portion of our test set. While it may provide a more focused prediction in some cases, this reduces the possibility of finding other related CWE to inform manual analysis when a match cannot be found among direct descendants. 

As shown in Figure ~\ref{fig:cwe-remaps}, the remaps happen within 1-4 steps of old CWE and this also includes the same parent's children. For example, in order to query for a better mapping to replace \emph{CWE-269: Improper Privilege Management}, which has no children in CWE-1003 view, we can query for its parent's (CWE-284) children, which are still closely related to CWE-269. Family candidates work best in our model, because they increase diversity while preserving relevance. Table ~\ref{tab:candidates_disc} summarizes the main advantages and limitations of these approaches.

\begin{table}[t]
\centering
\caption{\fixvw\ Discouraged Set Candidate Types with Advantages and Limitations}
\label{tab:candidates_disc}
\begin{tabular}{p{3.5cm}p{5.5cm}p{5.5cm}}
\toprule
\textbf{Candidate Set Type} & \textbf{Advantages} & \textbf{Limitations} \\
\midrule
\textbf{Descendants of old CWE} & Drives prediction toward more specific and valid weaknesses, preserves hierarchical relevance. & If the original mapping is incorrect, this can constrain the search to the wrong subtree. \\
\hline
\textbf{Family (Siblings)} & Explores closely related weaknesses under the same parent, increasing chance of capturing the correct mapping within a tight hierarchical scope. & May exclude correct CWEs if they lie in a different branch due to NVD misclassifications, too many candidates may confuse the model. \\

\bottomrule
\end{tabular}
\end{table}

\subsubsection{CWE Members for Views and Categories}

\begin{table}[h]
\centering
\caption{\fixvw\ Prohibited Set Candidate Types with Advantages and Limitations}
\label{tab:candidates_proh}
\begin{tabular}{p{3.5cm}p{5.5cm}p{5.5cm}}
\toprule
\textbf{Candidate Set Type} & \textbf{Advantages} & \textbf{Limitations} \\
\midrule
\textbf{MemberOf / HasMember Relationship} & Replaces invalid category mappings with valid members; preserves conceptual relevance. & Overly broad categories may introduce many low-relevance members, some categories are discontinued hence have no/few members. \\
\hline
\textbf{Find-Nearest-Neighbors (FNN)} & Includes structurally and ontologically related CWEs from KG links (\texttt{RelatedTo}), enabling recovery from some branch errors. & Since not many CVE are linked to these invalid CWE, ontology translation is not as informative as of those CWE that are valid and correctly mapped. \\
\hline
\end{tabular}
\end{table}%

For the Prohibited mappings, we use category and view members which carry conceptual relevance, while not as strong as hierarchical relevance. Besides the members, we also use descendants of these members, since the immediate members of the categories may be still overly broad Discouraged CWE, and using these as candidates is not an optimal strategy. Since members of a category do not necessarily come from the same branch of a hierarchical tree in the CWE database, the candidate sets may get overly large. On the contrary, some categories have none or very few members (cross section with CWE-1003 view) therefore we also tested another approach to populate the candidate lists. The Find Nearest Neighbors method of KG allows us to include ontologically related CWEs from KG links, allowing for an alternative to the default CWE-1003 view. The performance of this method is correlated to the centrality of the node, hence we may not achieve optimal performance with invalid CWE with very little connections to others. Table ~\ref{tab:candidates_proh} summarizes the main advantages and limitations of these approaches. 



\section{Implementation and Evaluation Methodology}\label{sec:methodology}

In this section, we detail the implementation and validation of \fixvw\, including data sets, knowledge graph construction and embedded model training. We then explain our methodology and metrics for evaluating the performance of \fixvw.


\subsection{Implementation Details}

Our implementation starts with the retrieval and processing of raw data from two different sources (NVD and MITRE). For CVE information and metadata, 
we obtain structured data provided by the NVD CVE API ~\cite{nvd-cve-api} that allows us to reach the latest CVE data along with the history of changes. Each record in this data includes fields such as CVE ID, description, affected platforms listed as CPE mappings, and root cause mappings in the form of CWE IDs. The history of changes include timestamps and complete modification information for each change, which allows users to filter by date and change type, and to access data added or removed during these modifications

The \emph{training dataset} corresponds to the NVD state on August 4, 2021, two years after the CWE-1003 view's finalization, allowing for subsequent corrections. The \emph{validation dataset} uses NVD data as of December 17, 2024, enabling evaluation of prediction accuracy on updated mappings.

The training dataset is used to train the knowledge graph, by converting mappings in the NVD (as provided in the CVE metadata) and CWE (as provided in the hierarchical views defining CWE relationships) to triple format.
The validation dataset serves the purpose of validating the embedding model's Top-ranked predictions, by predicting correct labels using our knowledge graph, and comparing our predictions with true labels updated after August 4, 2021 until December 17, 2024. Both datasets also include the CWE database (2021 version), including the CWE hierarchy of classes and views and their members.  

Between the 2021 and 2024 instances of the NVD, used for the training and validation datasets, we observe 334 Prohibited CWE mapping updates and 1467 Discouraged CWE mapping updates. Most of the Discouraged CWE mapping corrections were made as an effort to curate the CWE Top-25 lists of 2022, 2023 and 2024. As part of the Top-25 list curation process, the CWE team works together with CNAs on correctly mapping CVEs that are mapped to Class type CWEs, and update with a Base or Variant CWE whenever possible ~\cite{cwetop25} (see Figure ~\ref{fig:hierarchy} for CWE hierarchy detail levels). These CWEs are often Discouraged because they lack sufficient detail and have better suited, more accurate descendants (e.g., children). 




\subsubsection{Embedding Model Training Process}
For our embedding model, we adopt TransE~\cite{transE}, a widely used translational distance model, which has been shown to perform well~\cite{zpshitops} on sparse and hierarchical graphs similar to our vulnerability ontology. TransE operates under the principle that, for a valid triple $(h, r, t)$ (head, relation, tail), the embedding of the tail entity $t$ should be close to the embedding of the head entity $h$ translated by the embedding of the relation $r$.

To train the model, we use both \textit{positive triples} (valid facts from the KG) and \textit{negative triples} (artificially generated by corrupting one element of the triple). These negative examples are necessary to teach the model to discriminate valid relations from invalid ones. For example, a negative triple could be \texttt{(CVE-2020-17533, MatchingCWE, CWE-235)} since CWE-235 is unrelated to the CVE. To avoid overfitting and ensure numerical stability, our implementation uses the Adam optimizer ~\cite{kingma2017adam}, Multiclass Negative Log Likelihood (NLL) loss function~\cite{kadlec2017nll}, and an LP regularizer ~\cite{lacroix2018lp}  

\subsection{Implementation Details and Runtimes}
Model training, candidate set selection, and querying were all implemented in both PyTorch 2.6/Python 3.11 and AmpliGraph 1.4.0/Python 3.7.10.  Table~\ref{tab:gpu-cpu-comp} provides details on the computing facilities that were used to run our experiments. Table~\ref{tab:train-time-comp} shows a training time in the order of a few minutes for models run on GPUs. In the absence of GPUs, the training time of Ampligraph is shorter.

The time taken to determine the candidate set (Family method for Discouraged set and Member method for Prohibited set) and to query the Top-10 predictions for both methods is shown in Table \ref{tab:time-comp}. Notably, even when making predictions using the GPU, some operations (extracting Prohibited/Discouraged CWE and the CVEs mapped to them, determining the candidate set, and saving the results) still take place on the CPU for both implementations. As a result, there is an additional overhead for moving tensors from the GPU to CPU. For the PyTorch implementation, this could explain why the GPU execution time is actually slower than the CPU execution time - the time it takes to move tensors is greater than the speedup when using the GPU compared to the CPU. However, for AmpliGraph, the execution time improves because the time it takes to move tensors is shorter than the speedup when using the GPU compared to the CPU. 




\begin{table}[t]
    \centering
    \caption{GPU and CPU Specifications}
    \label{tab:gpu-cpu-comp}
    \begin{tabular}{|c|p{6cm}|p{6cm}|}
    \hline
         & BU Shared Computing Cluster (SCC) & New England Research Cloud (NERC) \\
         \hline
         GPU & 5120 CUDA cores, Tesla P100-PCIE-16GB & 3584 CUDA cores, Tesla V100-PCIE-32GB \\
         \hline
         CPU & 24 cores, Intel(R) Xeon(R) Gold 6132 CPU @ 2.60GHz 
             & 80 cores, Intel(R) Xeon(R) Gold 6148 CPU @ 2.40GHz \\
         \hline
    \end{tabular}
\end{table}

\begin{table}[t]
    \centering
    \caption{Comparison of training time on CPU and GPU for AmpliGraph and PyTorch implementations}
    \label{tab:train-time-comp}
    \begin{tabular}{|c|ccc|ccc|}
        \hline
        & \multicolumn{3}{c|}{\textbf{AmpliGraph 1.4.0}} & \multicolumn{3}{c|}{\textbf{PyTorch 2.6}} \\
        \cline{2-7}
        Device & \textit{Epochs }& \textit{Total Time} & \textit{Time/Epoch} & \textit{Epochs} & \textit{Total Time} & \textit{Time/Epoch} \\
        \hline
        \textbf{CPU} & 300  & 40:37 & 0:08.13 & 75 & 2:21:05 & 01:53.6 \\
        \hline
        \textbf{GPU} & 300 & 6:49 & 0:01.4 & 75 & 3:28 & 0:02.8 \\
        \hline
        
    \end{tabular}
\end{table}

\begin{table}[t]
\caption{Execution time comparison of \fixvw\ with AmpliGraph and PyTorch implementations on CPU (left) and GPU (right). There are a total of 338 Prohibited CVE and 1467 Discouraged CVE in the test set.} 
\label{tab:time-comp}
\centering
\begin{tabular}{ |c|c|c|c|c| } 
 \hline
    & \multicolumn{2}{c|}{\textbf{CPU}} & \multicolumn{2}{c|}{\textbf{GPU}} \\
    \cline{2-5}
    & AmpliGraph 1.4.0 (s) & PyTorch 2.6 (s) & AmpliGraph 1.4.0 (s) & PyTorch 2.6 (s)\\  
 \hline
 Prohibited Total & 385.59 & 0.148 & 289.717 & 0.2084 \\ 
 Prohibited Average & 1.144 & $4.39 \times 10^{-4}$  & 0.857 & $6.20 \times 10^{-4}$\\ 
 \hline
  Discouraged Total & 1367.263 &  0.674 & 1079.000 &  1.133 \\ 
 Discouraged Average & 0.932 & $4.39 \times 10^{-4}$ & 0.736 & $7.72 \times 10^{-4}$\\ 
 \hline
\end{tabular}

\end{table}

\subsection{Evaluation Methodology}
\label{sec:eval-method}
We evaluate our trained embedding model, by using \emph{rank-based metrics}, which assess how well the model can predict the correct entity in a triple when given partial information, and by using {coverage metrics} to evaluate whether one of the Top-10 predictions provides an exact match, a fine grain match or a coarse grain match.
The evaluation algorithm is provided by Algorithm~\ref{alg:eva}.

\begin{algorithm}[!t]
    \caption{Evaluate-FixV2W}\label{alg:eva}
    \begin{algorithmic}[1]
        \State  Determine the test set $V^T = \{v \in V\}$, 
        such that $v$ is mapped to a Prohibited CWE $w_1$ in the training dataset and to an allowed CWE $w_2$ in the validation dataset. 
        \State Run procedure \textproc{FixV2W($V^T$)}
        \ForAll{$v \in V^T$}:
            \State Scan the sorted candidate list of weaknesses $w \in W^A(v)$, starting from the first element
            \If{$\exists w$ such that $w=w_2$} 
                \State Exact match found
                \State Return rank of $w$ 
            \EndIf 
           \If{$\exists w$ such that $w$ is a  direct neighbor of $w_2$} 
                \State Fine grain match found
                \State Return rank of $w$ 
            \EndIf 
          \If{$\exists w$ such that $w$ is in the same branch as $w_2$} 
                \State Coarse grain match found
                \State Return rank of $w$  
          \EndIf      
       \EndFor  \\       
\Return{Accuracy, MR and MRR of exact matches, fine grain matches, and coarse grain matches}
    \end{algorithmic}
\end{algorithm}

\subsubsection{Rank-based metrics}
We apply ranking metrics commonly used in conjunction with knowledge graphs. The
\textit{Mean Rank (MR)} computes the average rank position of the correct entity across all test queries. A lower MR indicates better performance. The \textit{Mean Reciprocal Rank (MRR)} computes the average of the reciprocal ranks (1/rank) for the correct entity. MRR emphasizes the importance of placing the correct entity as high as possible in the ranking. An MRR of 1.0 means perfect ranking (all correct answers are ranked first).
Last, the \textit{Hits@N} metric measures the proportion of times the correct entity appears in the Top-$N$ predicted results. In our evaluations, we use $N=1,3,5,10$ and $20$.

\subsubsection{Coverage Metrics}

We evaluate how much of the test instances (triples) are correctly predicted by our model among the Top-10 ranked predictions. We recall that the \fixvw\ method, described in Algorithm~\ref{alg:cap}, returns a set of candidates $W^A(v)$ ranked in sorted order from the most likely  the least. In our evaluation, if \fixvw\ can predict the exact true triple within the Top-10 positions, we call this an \textit{exact match.} If the model can predict a neighbor of the correct label, we call this a \textit{fine-grain match}. As discussed in Section~\ref{sec:more-accurate}, we empirically found several cases where the fine-grain match provided by \fixvw\ seems more precise that the the (corrected) NVD label used as the ground-truth. Last, a prediction is referred to as a \textit{coarse-grain match}, if the predicted CWE belongs to the same branch as the correct CWE (e.g., descendant of the same parent CWE) . 

To illustrate these concepts, consider the example shown in Figure ~\ref{fig:hierarchy} (see Section~\ref{sec:background}). Suppose the true label is \texttt{CWE-140} and the model returns \texttt{\{CWE-228, CWE-138, CWE-707, CWE-140\}} as its Top-rankings. The result implies an exact match at rank 4, a fine-grain match at rank 2, and coarse-grain matches at ranks 1 and 3.

\section{Experimental Results}\label{sec:eval}



In this section, we present our experimental results and key findings. 
First, we evaluate the performance of \fixvw\ on correcting Discouraged and Prohibited mappings.
Next, we show how our fixes to the NVD help improve ML models relying on this database, specifically models for uncovering unknown relationships between CVE and CWE entries.
Last, we demonstrate how our model could mitigate exploitation of CVEs initially mapped to erroneous CWEs.   

\subsection{Performance Evaluation of FixV2W}

In the 2021 NVD version used for the the training dataset, we found 13,784 Prohibited CVE  to CWE mappings. Among those, 336 (i.e, 2.5\%) were remapped to allowed mappings in the validation dataset of December 2024. Likewise, we found 33,928 Discouraged CVE-CWE mappings in the training dataset. Among those, 1457 (i.e., 4.3\%) were remapped to allowed mappings in the validation dataset. 
We work with labeled test sets that include these updated CWE mappings as the true labels, and evaluate the performance of \fixvw using the evaluation methodology described in Section~\ref{sec:eval-method}. 


\subsubsection{Evaluation of FixV2W for Correcting Discouraged CWEs}


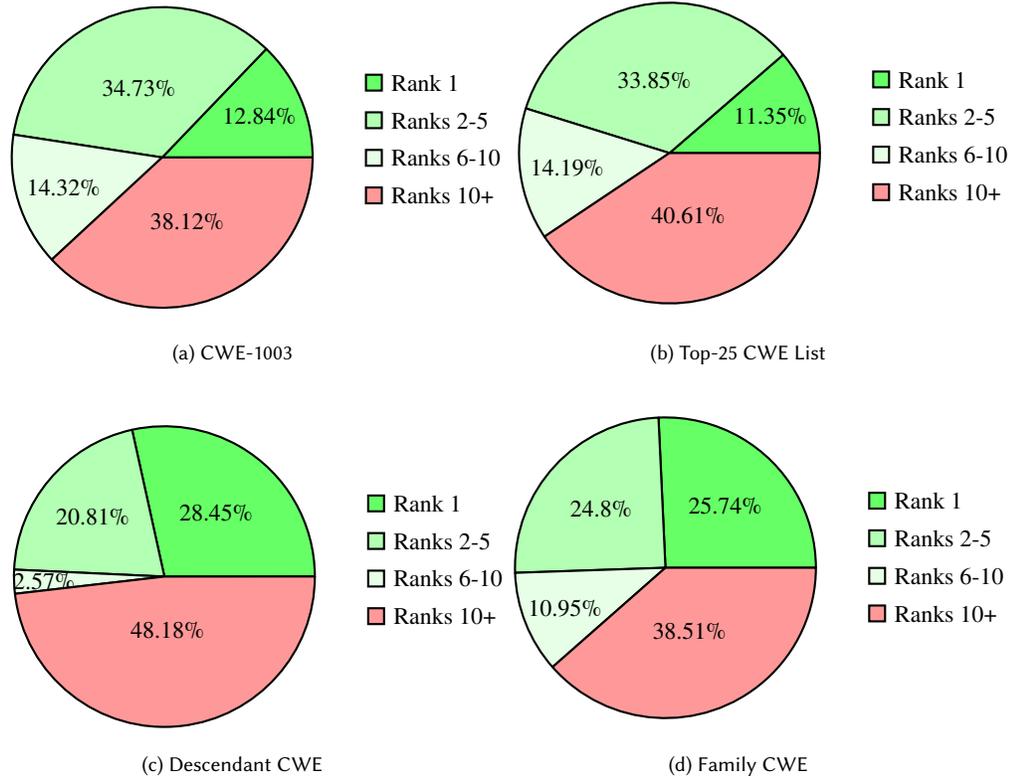
\begin{figure}[t]
    \centering
    \begin{subfigure}{0.4\linewidth}
        \centering
        \begin{tikzpicture}
        \pie[text=inside, radius=2, text = legend, color={green!60, green!30, green!10, red!40}]{12.84/Rank 1, 34.73/Ranks 2-5, 14.32/Ranks 6-10, 38.12/Ranks 10+}
        \end{tikzpicture}
        \caption{CWE-1003}
        \label{fig:cwe-1003-d}
    \end{subfigure}%
    \hspace{6mm}
    \begin{subfigure}{0.4\linewidth}
        \centering
        \begin{tikzpicture}
        \pie[text=inside, radius=2, text=legend, color={green!60, green!30, green!10, red!40}]{11.35/Rank 1, 33.85/Ranks 2-5, 14.19/Ranks 6-10, 40.61/Ranks 10+}
        \end{tikzpicture}
        \caption{Top-25 CWE List}
        \label{fig:cwe-25-d}
    \end{subfigure}

    \vspace{6mm}

    \begin{subfigure}{0.4\linewidth}
        \centering
        \begin{tikzpicture}
        \pie[color={green!60, green!30, green!10, red!40},text=pin,  text=legend,  radius=2,]{28.45/Rank 1,  20.81/Ranks 2-5,  2.57/Ranks 6-10,  48.18/Ranks 10+}
        \end{tikzpicture}
        \caption{Descendant CWE}
        \label{fig:cwe-child-d}
    \end{subfigure}%
    \hspace{6mm}
    \begin{subfigure}{0.4\linewidth}
        \centering
        \begin{tikzpicture}
        \pie[color={green!60, green!30, green!10, red!40},text=inside, radius=2, text=legend]{25.74/Rank 1, 24.8/Ranks 2-5, 10.95/Ranks 6-10, 38.51/Ranks 10+}
        \end{tikzpicture}
        \caption{Family CWE}
        \label{fig:cwe-ext-d}
    \end{subfigure}

    \caption{Distribution of ranks for CVE-CWE prediction of Discouraged mappings using different candidate sets.}
    \label{fig:cwe-pie-d}
\end{figure}


Figure~\ref{fig:cwe-pie-d} summarizes the performance of \fixvw\ for different CWE candidate lists.
As a baseline, we use the CWE-1003 view as the entity candidate list, for each CVE in the test set regardless of their related CWEs. 
First, we see that 61.88\% of correct matches were among the Top-10 positions. The Top-1 match percentage, however, is quite low at 12.84\% and inadequate for performing the automated correction task on these predictions alone.

The second candidate set is the CWE Top-25 view. This view includes the most commonly mapped CWE in 2021 (recall that the 2021 dataset is used for training). Since most CVEs were remapped to these CWE, this approach achieves good coverage performance overall. The main issue is that it focuses too much on highly central CWEs and ignore less popular, but still relevant CWEs. In fact, the Top-1 match percentage is even lower than for the CWE-1003, at 11.35\%, because is it highly affected by node centrality. Thus, popular CWEs, such as CWE-20 and CWE-119, appear at the Top-rank no matter which CVE we query for.  

Since Discouraged CWEs are often higher up in the hierarchy, based on our longitudinal analysis, a reasonable method is to replace them with one of their \emph{descendants}.
For this candidate set, we select  CWEs by applying the \texttt{ChildOf} relationship in the CWE hierarchy. Only those included in the CWE-1003 view are taken into consideration, to conform with the NVD requirements. We finds that the overall Top-N match percentages are the best for this candidate set and 48.26\% of the predictions appear in the Top-5 positions, and only 2.75\% appear between ranks 6 and 10.
Although the model's rank based metrics perform best with this candidate set, the coverage metrics are poorer. While increasing focus, this method sometimes limits the number of candidates to as little as one or two. When the descendant CWEs are not exact matches, there are no candidates left that could even be fine-grain or coarse-grain matches. This especially happens if the original mapping was the wrong child of a correct class CWE. 


Alternatively, when there are fewer than 10 descendants to replace a Discouraged CWE, we consider an extended  candidate entity set, which we refer to as \emph{family}. In this case, when possible, we take one step up in the hierarchy (i.e., applying the \texttt{ParentOf} relationship on the old CWE), and populate the candidate entity list with the descendants of that node (limiting ourselves as usual to CWEs belonging to the CWE-1003 view). This ensures that we have a more comprehensive list of candidates for the \fixvw algorithm. 
Figure~\ref{fig:cwe-pie-d}(c) shows that this approach reaches almost the same overall coverage as CWE-1003 baseline, with an increased Top-rank coverage, climbing from 12.84\% to 25.74\%. Thus, we can argue that the extended family candidate set performs best, even though its Top-N match percentages seem slightly lower than for the descendant list. Overall, the number of CVE-CWE mappings predicted among Top-1, Top-5 and Top-10 positions do increase. 

\subsubsection{Evaluation of FixV2W for Correcting Prohibited CWEs}
\begin{figure}[t]
    \centering
    \begin{subfigure}{0.4\linewidth}
        \centering
        \begin{tikzpicture}
       \pie[text=inside, radius=2, text=legend, color={green!60, green!30, green!10, red!40}]{9.44/Rank 1, 17.40/Ranks 2-5, 10.32/Ranks 6-10, 62.83/Ranks 10+}
        \end{tikzpicture}
        \caption{CWE-1003}
        \label{fig:cwe-1003-p}
    \end{subfigure}%
    \hspace{6mm}
    \begin{subfigure}{0.40\linewidth}
        \centering
        \begin{tikzpicture}
        \pie[text=inside, radius=2, text=legend, color={green!60, green!30, green!10, red!40}]{7.08/Rank 1, 13.86/Ranks 2-5, 17.109/Ranks 6-10, 61.95/Ranks 10+}
        \end{tikzpicture}
        \caption{Top-25 CWE List 2021}
        \label{fig:cwe-25-p}
    \end{subfigure}
    \vspace{6mm}
    \begin{subfigure}{0.40\linewidth}
        \centering
        \begin{tikzpicture}
        \pie[text=inside, radius=2, text=legend, color={green!60, green!30, green!10, red!40}]{15.93/Rank 1, 33.63/Ranks 2-5, 13.86/Ranks 6-10, 36.58/Ranks 10+}
        \end{tikzpicture}
        \caption{Member CWE}
        \label{fig:cwe-mem-p}
    \end{subfigure}%
    \hspace{6mm}
    \begin{subfigure}{0.40\linewidth}
        \centering
        \begin{tikzpicture}
        \pie[text=inside, radius=2, text = legend, color={green!60, green!30, green!10, red!40}]{15.63/Rank 1, 32.74/Ranks 2-5, 15.63/Ranks 6-10, 35.99/Ranks 10+}
        \end{tikzpicture}
        \caption{Nearest Neighbor Optimization}
        \label{fig:cwe-nn-p}
    \end{subfigure}
    \caption{Distribution of ranks for CVE-CWE prediction of Prohibited mappings using different candidate sets.}
    \label{fig:cwe-pie-p}
\end{figure}
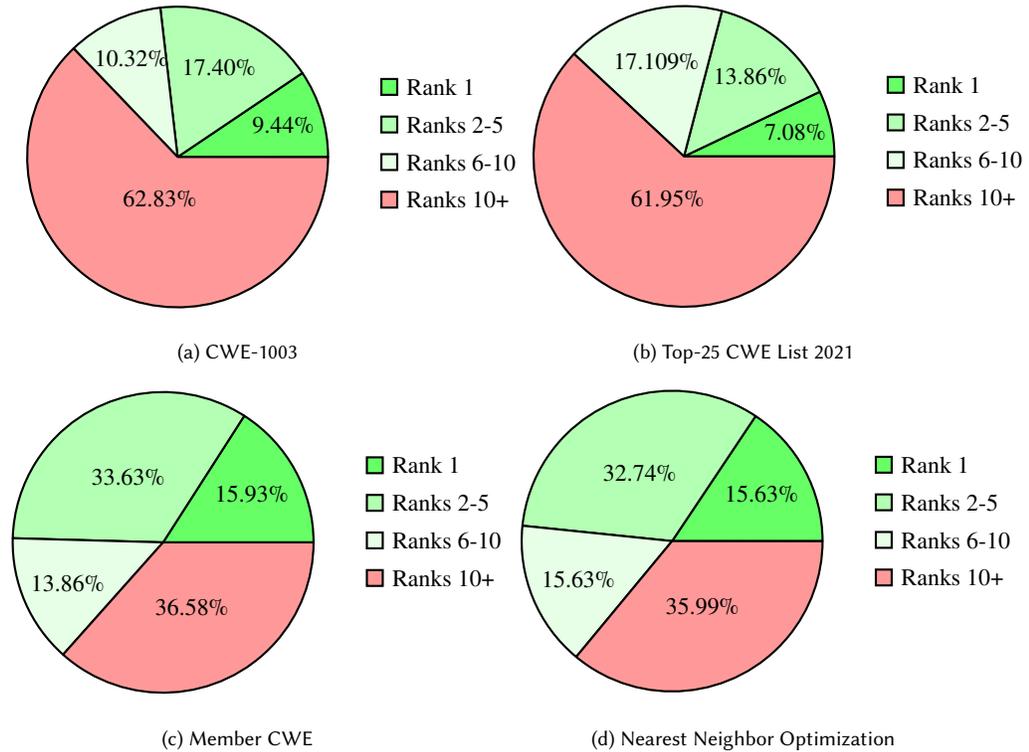

Figure~\ref{fig:cwe-pie-p} summarizes the performance of \fixvw\ for correcting Prohibited mappings. As for Discouraged mappings, we consider different CWE candidate lists. Beyond the CWE-1003 and Top-25 baselines, we consider a list of the \emph{members} of the category or view corresponding to the prior Prohibited CWE.  
For each Prohibited CWE, which is a category or view, our method involves extracting the \texttt{MemberOf} relationship from the CWE database. The member candidate entity sets provides us with the most related CWEs for each Prohibited CWE. This choice is again driven by the findings of our longitudinal study. 

Similar to Discouraged mappings, we also consider an extended set that includes a larger number of candidates, which we refer to as \emph{find nearest neighbors (FNN)} list to include other CWEs that are in close proximity in the KG embedding vector space. FNN is a knowledge graph method to extract neighbors, that are most likely associated with a node, even if there is no direct relationship between the two nodes. This method uses the vector space to calculate distances from a node to a list on candidates, and returns the Top-K nearest nodes. 


Our experimental results indicate that the Member CWE candidate list significantly outperforms the CWE-1003 and Top-25 baselines, both in terms of coverage and precision. Figure~\ref{fig:cwe-mem-p} shows the increase in coverage and the proportion of correct labels predicted in Top-1, Top-5 and Top-10 ranks. We achieve an exact match coverage that is twice as much (137 out of 338) that of  the Top-25 list (67 out of 338), and the total correct prediction count, with either exact match, fine grain match or coarse grain match, goes up from 135 to 215. The Nearest Neighbor Optimization (FNN) candidate list performs comparably to the Member CWE approach.

\subsection{Coverage vs. Accuracy}
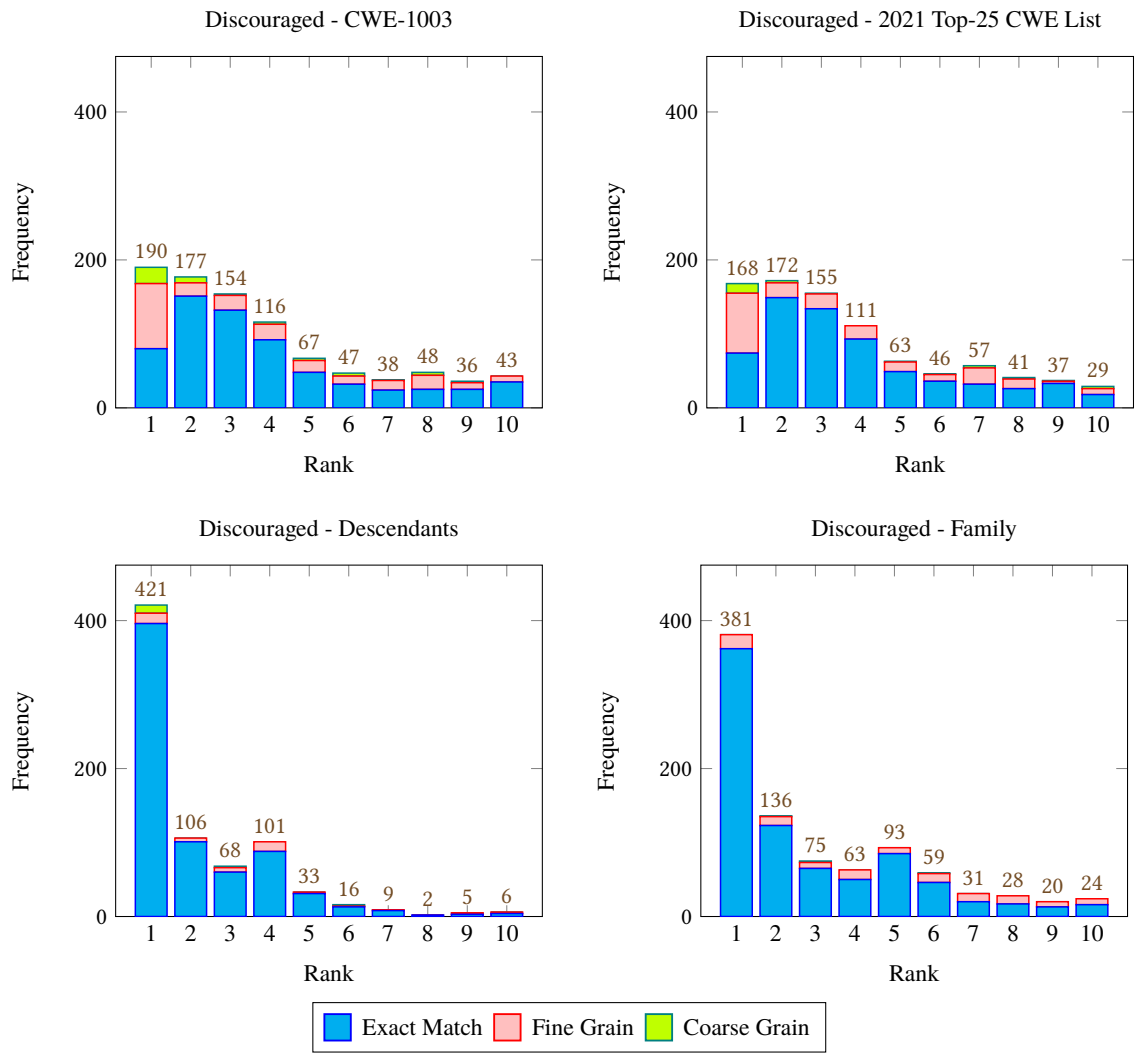
\begin{figure} []
\begin{minipage}{0.48\textwidth}
\centering
\begin{tikzpicture}
\begin{axis}[
    ybar stacked,
    bar width=12pt,
    width=\textwidth,
    ylabel={Frequency},
    ymax = 475,
    xlabel={Rank},
    xtick=data,
    xticklabels from table={\Discouragedthree}{Rank},
    ymin=0,
    legend style={at={(0.5,-0.25)}, anchor=north, legend columns=1},
    title={Discouraged - Descendants}
]

\addplot+[draw = blue,line width = .2mm,fill=cyan] table[x expr=\coordindex, y=Exact] {\Discouragedthree};
\addplot+[draw = red,line width = .2mm,fill = pink] table[x expr=\coordindex, y=Fine] {\Discouragedthree};
\addplot+[draw = teal,line width = .2mm, fill=lime, nodes near coords, 
        point meta=y,
        nodes near coords align={anchor=south}] table[x expr=\coordindex, y=Coarse] {\Discouragedthree};


\end{axis}
\end{tikzpicture}
\end{minipage}
\hfill
\begin{minipage}{0.48\textwidth}
\centering
\begin{tikzpicture}
\begin{axis}[
    ybar stacked,
    bar width=12pt,
    width=\textwidth,
    xlabel={Rank},
    ylabel={Frequency},
    ymax = 475,
    xtick=data,
    xticklabels from table={\Discouragedfour}{Rank},
    ymin=0,
    legend style={at={(0.5,-0.25)}, anchor=north, legend columns=3},
    title={Discouraged - Family}
]

\addplot+[draw = blue,line width = .2mm,fill=cyan] table[x expr=\coordindex, y=Exact] {\Discouragedfour};
\addplot+[draw = red,line width = .2mm,fill = pink] table[x expr=\coordindex, y=Fine] {\Discouragedfour};
\addplot+[draw = teal,line width = .2mm, fill=lime, nodes near coords, 
        point meta=y,
        nodes near coords align={anchor=south}] table[x expr=\coordindex, y=Coarse] {\Discouragedfour};


\end{axis}
\end{tikzpicture}
\end{minipage}
\vspace{0.6mm}
\begin{center}
\begin{tikzpicture}
\node[draw, inner sep=2pt] {
    \begin{tikzpicture}
\matrix [matrix of nodes, nodes={anchor=west}, column sep=2pt] {
    \node[draw=blue, fill=cyan, line width=.25mm, minimum width=10pt, minimum height=10pt]{}; & \node[]{Exact Match}; &
    \node[draw=red, fill=pink, line width=.25mm, minimum width=10pt, minimum height=10pt]{}; & \node[]{Fine Grain}; &
    \node[draw=teal, fill=lime, line width=.25mm, minimum width=10pt, minimum height=10pt]{}; & \node[]{Coarse Grain }; \\
};
\end{tikzpicture}
 };
\end{tikzpicture}
\end{center}
\caption{Discouraged set predictions rank distributions for each candidate set.}
\label{fig:dist-disc}
\end{figure}

\begin{figure}[t]
\centering

\begin{minipage}{0.48\textwidth}
\centering
\begin{tikzpicture}
\begin{axis}[
    ybar stacked,
    bar width=12pt,
    width=\textwidth,
    ylabel={Frequency},
    ymax = 70,
    xlabel={Rank},
    xtick=data,
    xticklabels from table={\Prohibitedthree}{Rank},
    ymin=0,
    legend style={at={(0.5,-0.25)}, anchor=north, legend columns=1},
    title={Prohibited - Members}
]
\addplot+[draw = blue,line width = .2mm,fill=cyan] table[x expr=\coordindex, y=Exact] {\Prohibitedthree};
\addplot+[draw = red,line width = .2mm,fill = pink] table[x expr=\coordindex, y=Fine] {\Prohibitedthree};
\addplot+[draw = teal,line width = .2mm, fill=lime, nodes near coords, 
        point meta=y,
        nodes near coords align={anchor=south}] table[x expr=\coordindex, y=Coarse] {\Prohibitedthree};


\end{axis}
\end{tikzpicture}
\end{minipage}
\hfill
\begin{minipage}{0.48\textwidth}
\centering
\begin{tikzpicture}
\begin{axis}[
    ybar stacked,
    bar width=12pt,
    width=\textwidth,
    ymax = 70,
    xlabel={Rank},
    ylabel={Frequency},
    xtick=data,
    xticklabels from table={\Prohibitedfour}{Rank},
    ymin=0,
    legend style={at={(0.5,-0.25)}, anchor=north, legend columns=3},
    title={Prohibited - Nearest Neighbors}
]
\addplot+[draw = blue,line width = .2mm,fill=cyan] table[x expr=\coordindex, y=Exact] {\Prohibitedfour};
\addplot+[draw = red,line width = .2mm,fill = pink] table[x expr=\coordindex, y=Fine] {\Prohibitedfour};
\addplot+[draw = teal,line width = .2mm, fill=lime, nodes near coords, 
        point meta=y,
        nodes near coords align={anchor=south}] table[x expr=\coordindex, y=Coarse] {\Prohibitedfour};
\end{axis}
\end{tikzpicture}
\end{minipage}
\vspace{0.6mm}
\begin{center}
\begin{tikzpicture}
\node[draw, inner sep=2pt] {
    \begin{tikzpicture}
\matrix [matrix of nodes, nodes={anchor=west}, column sep=2pt] {
    \node[draw=blue, fill=cyan, line width=.25mm, minimum width=10pt, minimum height=10pt]{}; & \node[]{Exact Match}; &
    \node[draw=red, fill=pink, line width=.25mm, minimum width=10pt, minimum height=10pt]{}; & \node[]{Fine Grain}; &
    \node[draw=teal, fill=lime, line width=.25mm, minimum width=10pt, minimum height=10pt]{}; & \node[]{Coarse Grain }; \\
};
\end{tikzpicture}
 };
\end{tikzpicture}
\end{center}
\caption{Prohibited set predictions rank distributions for each candidate set.}
\label{fig:dist-proh}
\end{figure}
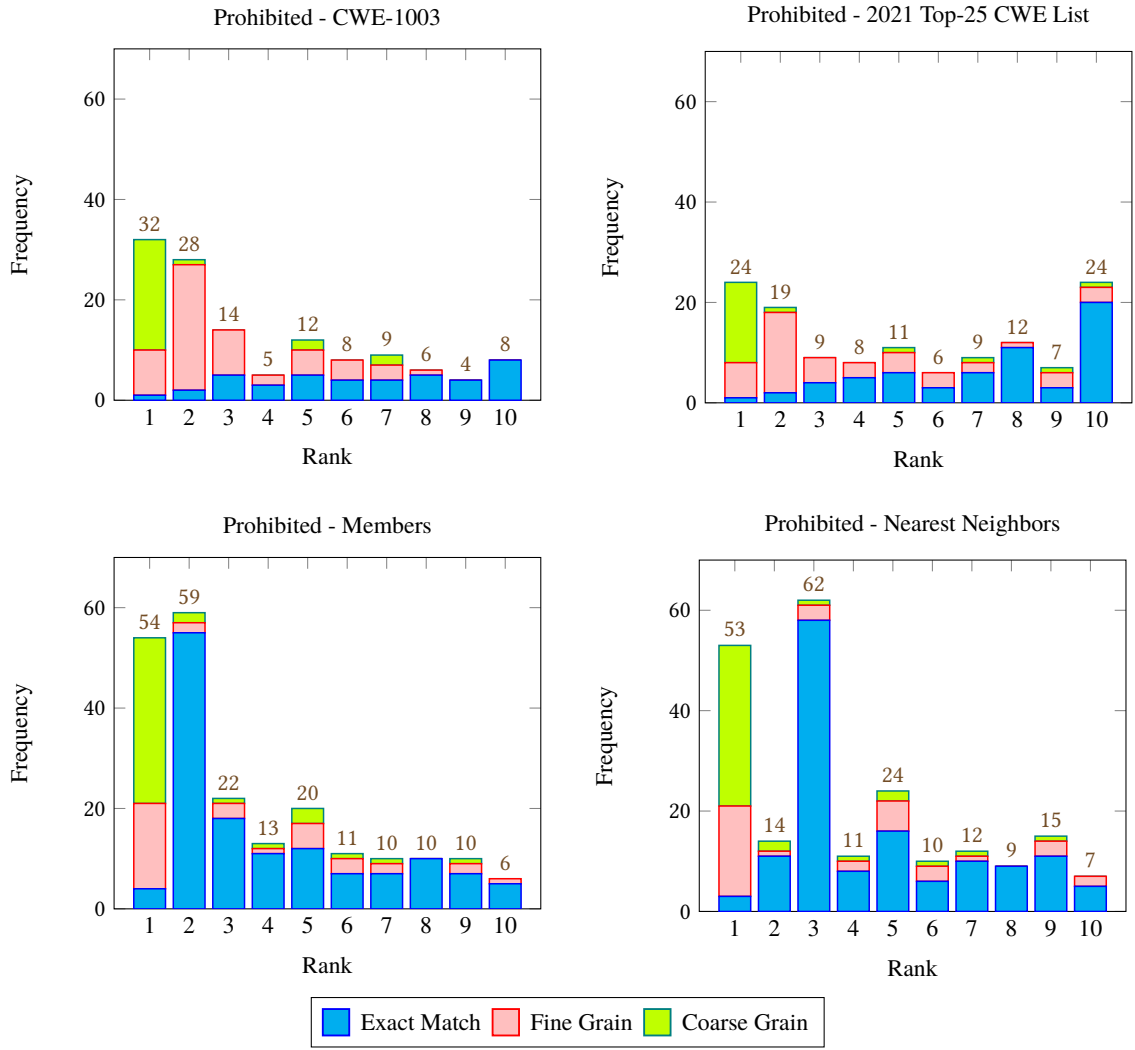

Figures~\ref{fig:dist-disc} and ~\ref{fig:dist-proh} display the rank distribution for each of the candidate sets. Focusing on Figure \ref{fig:dist-disc}, when using the CWE-1003 view, we have a large candidate set and therefore fewer results outside the Top-10. However, the distribution of ranks is scattered since \fixvw\ returns correct predictions pretty much uniformly across all ranks.
However, with smaller candidate sets (i.e., Descendants and Family), \fixvw\ is able to predict many more CWEs at the Top-ranks, thanks to focused predictions. Furthermore, most of the predictions are exact matches or fine grain matches, which demonstrates the model's ability to accurately predict CWEs, requiring little to no manual corrections afterwards.

Focusing only on the exact matches (blue bars) in Figure~\ref{fig:dist-proh}, we see that Top-25 CWE candidate set performs better than the CWE-1003 baseline, with fewer Top-rank matches. On the bottom, Member CWE and  Member + Nearest Neighbors candidate entity set rank distributions are shown. We see that although the total number of predicted true labels are more or less the same, 
using only member CWE performs better with more correct CWE predicted at rank 2, rather than supplementing with Nearest Neighbors which results in more correct predictions on rank 3. Overall, since using nearest neighbors to fill up the candidate lists provides no apparent benefit, the best approach for correcting Prohibited CWE mappings is by using member CWE. We discuss the reason of this trend in more detail in section ~\ref{sec:incorrect}.

\pgfplotstableread{
Rank  Exact  Fine  Coarse  Overall
1   1      9     22      32
2   2      25    1       28
3   5      9     0       14
4   3      2     0       5
5   5      5     2       12
6   4      4     0       8
7   4      3     2       9
8   5      1     0       6
9   4      0     0       4
10  8      0     0       8
}\Prohibitedone

\pgfplotstableread{
Rank  Exact  Fine  Coarse  Overall
1 80 88 22 190
2 151 18 8 177
3 132 20 2 154
4 92 21 3 116
5 48 16 3 67
6 32 11 4 47
7 24 13 1 38
8 25 19 4 48
9 25 9 2 36
10 35 8 0 43
}\Discouragedone

\pgfplotstableread{
Rank  Exact  Fine  Coarse  Overall
1 1 7 16 24  
2 2 16 1 19  
3 4 5 0 9  
4 5 3 0 8  
5 6 4 1 11  
6 3 3 0 6  
7 6 2 1 9  
8 11 1 0 12  
9 3 3 1 7  
10 20 3 1 24
}\Prohibitedtwo

\pgfplotstableread{
Rank  Exact  Fine  Coarse  Overall
1 74 81 13 168  
2 149 20 3 172  
3 134 20 1 155  
4 93 18 0 111  
5 49 13 1 63  
6 36 9 1 46  
7 32 22 3 57  
8 26 13 2 41  
9 33 3 1 37  
10 18 8 3 29
}\Discouragedtwo

\pgfplotstableread{
Rank  Exact  Fine  Coarse  Overall
1 4 17 33 54  
2 55 2 2 59  
3 18 3 1 22  
4 11 1 1 13  
5 12 5 3 20  
6 7 3 1 11  
7 7 2 1 10  
8 10 0 0 10  
9 7 2 1 10  
10 5 1 0 6
}\Prohibitedthree

\pgfplotstableread{
Rank  Exact  Fine  Coarse  Overall
1 396 14 11 421  
2 101 5 0 106  
3 60 6 2 68  
4 88 13 0 101  
5 31 2 0 33  
6 13 1 2 16  
7 8 1 0 9  
8 2 0 0 2  
9 3 2 0 5  
10 4 2 0 6
}\Discouragedthree

\pgfplotstableread{
Rank  Exact  Fine  Coarse  Overall
1 3 18 32 53  
2 11 1 2 14  
3 58 3 1 62  
4 8 2 1 11  
5 16 6 2 24  
6 6 3 1 10  
7 10 1 1 12  
8 9 0 0 9  
9 11 3 1 15  
10 5 2 0 7
}\Prohibitedfour

\pgfplotstableread{
Rank  Exact  Fine  Coarse  Overall
1 362 19 0 381  
2 123 12 1 136  
3 65 8 2 75  
4 50 13 0 63  
5 85 8 0 93  
6 46 12 1 59  
7 20 11 0 31  
8 17 11 0 28  
9 13 7 0 20  
10 16 8 0 24
}\Discouragedfour

\pgfplotstableread{
Rank  Exact  Fine  Coarse  Overall
1 52 9 2 63  
2 14 12 0 26  
3 13 7 0 20  
4 11 2 0 13  
5 11 6 1 18  
6 3 3 0 6  
7 2 2 0 4  
8 0 0 0 0  
9 0 0 0 0  
10 1 0 0 1  
}\Prohibitedchild

\begin{figure} []
\begin{minipage}{0.48\textwidth}
\centering
\begin{tikzpicture}
\begin{axis}[
    ybar stacked,
    bar width=12pt,
    width=\textwidth,
    ylabel={Frequency},
    ymax = 475,
    xlabel={Rank},
    ylabel={Frequency},
    xtick=data,
    xticklabels from table={\Discouragedone}{Rank},
    ymin=0,
    legend style={at={(0.5,-0.25)}, anchor=north, legend columns=1},
    title={Discouraged - CWE-1003}
]

\addplot+[draw = blue,line width = .2mm,fill=cyan] table[x expr=\coordindex, y=Exact] {\Discouragedone};
\addplot+[draw = red,line width = .2mm,fill = pink] table[x expr=\coordindex, y=Fine] {\Discouragedone};
\addplot+[draw = teal,line width = .2mm, fill=lime, nodes near coords, 
        point meta=y,
        nodes near coords align={anchor=south}] table[x expr=\coordindex, y=Coarse] {\Discouragedone};


\end{axis}
\end{tikzpicture}
\end{minipage}
\hfill
\begin{minipage}{0.48\textwidth}
\centering
\begin{tikzpicture}
\begin{axis}[
    ybar stacked,
    bar width=12pt,
    width=\textwidth,
    xlabel={Rank},
    ylabel={Frequency},
    ymax = 475,
    xtick=data,
    xticklabels from table={\Discouragedtwo}{Rank},
    ymin=0,
    legend style={at={(0.5,-0.25)}, anchor=north, legend columns=1},
    title={Discouraged - 2021 Top-25 CWE List}
]

\addplot+[draw = blue,line width = .2mm,fill=cyan] table[x expr=\coordindex, y=Exact] {\Discouragedtwo};
\addplot+[draw = red,line width = .2mm,fill = pink] table[x expr=\coordindex, y=Fine] {\Discouragedtwo};
\addplot+[draw = teal,line width = .2mm, fill=lime, nodes near coords, 
        point meta=y,
        nodes near coords align={anchor=south}] table[x expr=\coordindex, y=Coarse] {\Discouragedtwo};


\end{axis}
\end{tikzpicture}
\end{minipage}
\end{figure}

\begin{figure}[t]
\centering
\begin{minipage}{0.48\textwidth}
\centering
\begin{tikzpicture}
\begin{axis}[
    ybar stacked,
    bar width=12pt,
    width=\textwidth,
    ylabel={Frequency},
    ymax = 70,
    xlabel={Rank},
    xtick=data,
    xticklabels from table={\Prohibitedone}{Rank},
    ymin=0,
    legend style={at={(0.5,-0.25)}, anchor=north, legend columns=1},
    title={Prohibited - CWE-1003}
]

\addplot+[draw = blue,line width = .2mm, fill=cyan] table[x expr=\coordindex, y=Exact] {\Prohibitedone};
\addplot+[draw = red,line width = .2mm, fill = pink] table[x expr=\coordindex, y=Fine] {\Prohibitedone};
\addplot+[draw = teal,line width = .2mm, fill=lime, nodes near coords, 
        point meta=y,
        nodes near coords align={anchor=south}] table[x expr=\coordindex, y=Coarse] {\Prohibitedone};

\end{axis}
\end{tikzpicture}
\end{minipage}
\hfill
\begin{minipage}{0.48\textwidth}
\centering
\begin{tikzpicture}
\begin{axis}[
    ybar stacked,
    bar width=12pt,
    width=\textwidth,
    ylabel={Frequency},
    ymax = 70,
    xlabel={Rank},
    xtick=data,
    xticklabels from table={\Prohibitedtwo}{Rank},
    ymin=0,
    legend style={at={(0.5,-0.25)}, anchor=north, legend columns=1},
    title={Prohibited - 2021 Top-25 CWE List}
]

\addplot+[draw = blue,line width = .2mm,fill=cyan] table[x expr=\coordindex, y=Exact] {\Prohibitedtwo};
\addplot+[draw = red,line width = .2mm,fill = pink] table[x expr=\coordindex, y=Fine] {\Prohibitedtwo};
\addplot+[draw = teal,line width = .2mm, fill=lime, nodes near coords, 
        point meta=y,
        nodes near coords align={anchor=south}] table[x expr=\coordindex, y=Coarse] {\Prohibitedtwo};


\end{axis}
\end{tikzpicture}
\end{minipage}
\end{figure}

\subsection{Uncovering Unknown Associations between CVE and CWE Entries}

Correcting Prohibited and Discouraged CVE-CWE mappings is important both for increasing accuracy of the NVD, and for improving ML models relying on the NVD. Specifically, we show next how the updated NVD dataset, obtained after applying  \fixvw\ on the original database, helps improving the performance of knowledge graph completion. The task of knowledge graph completion involves inferring missing triples within this graph (in our case, missing associations in the NVD).


The work in~\cite{zpshitops} uses a graph completion approach to predict unknown associations between CVE and CWE entries, leveraging the NVD database. Here, we apply the same method, but 
use the CVE-CWE mappings returned by \fixvw\ to replace the Prohibited and Discouraged CWEs in the NVD. For Discouraged CWEs, we use the \emph{family} candidate entity set. For the Prohibited CWEs, we used members (and their descendants). We evaluate the performance of the model when replacing Discouraged/Prohibited mappings by either the Top-1 match returned by \fixvw\ or the Top-2 or Top-3 matches.

We compare the performance of the model of~\cite{zpshitops} with our updated model on predicting hidden CVE-CWE mappings (currently unknown, but revealed in the future), using the original August 2021 NVD dataset in the former case and the updated August 2021 NVD dataset (after applying \fixvw) in the latter case. 



\begin{table}[t]
    \centering
    \caption{Open-world evaluation of graph completion, comparing the performance of our model, based on the updated NVD after applying \fixvw, and that of~\cite{zpshitops}, based on the original NVD. The model was trained on NVD as of August 2021 and tested on new triples added to the NVD between August 2021 and November 2021.}
    \label{tab:openworldretrained}
    \vspace{1em} 
    \begin{tabular}{|c||c|c|c|c|}
    \hline
    \textbf{CVE-CWE} & \textbf{Original~\cite{zpshitops}} & \textbf{Fixed (Top-1)} & \textbf{Fixed (Top-2)} & \textbf{Fixed (Top-3)} \\
    \hline
    MRR & 0.143  & 0.261 & \textbf{0.608} & 0.573\\
    MR & 38 &  24.719 & \textbf{14.711} & 15.053\\
    Hits@20 & 0.500  &  0.717 & \textbf{0.876}   & 0.848\\
    Hits@10 & 0.358 &  0.504 & \textbf{0.793}   & 0.744\\
    Hits@3 & 0.118  & 0.249  & \textbf{0.640} & 0.616 \\
    Hits@1 & 0.059 & 0.160 & \textbf{0.521} & 0.479  \\
    \hline
    \end{tabular}
\end{table}

Table~\ref{tab:openworldretrained} shows the performance of the models under an open-world assumption (OWA) which is generally considered more realistic. 
We observe that the updated model achieves much better performance. Interestingly, the best performance is achieved by replacing each discouraged/prohibited mapping by the Top-2 matches returned by \fixvw, improving the MRR score from 0.173 to 0.608 and the Hits@10 score from 0.389 to 0.793, compared to the model of~\cite{zpshitops} . This  is because \fixvw\ often predicts the correct match at the second rank, as shown previously. Moreover, due to the multi-label classification nature of CVEs, most of the vulnerabilities are caused by several, often chained, weaknesses. 

Table~\ref{tab:closedworldretrained} shows the performance of the models under a closed-world assumption This splits the KG into training and validation sets both formed from the August 2021 NVD dataset. In this case, the Top-3 model performs the best, showing a significant jump in the Mean Reciprocal Rank (MRR) score from 0.419 to 0.731, which indicates more concentrated correct predictions at top ranks. 
Overall, we observe that the performance of the graph completion model improves significantly after updating the Prohibited and Discouraged CVE-CWE mappings, using \fixvw. 

\begin{table}[t]
\centering
\caption{Closed-world evaluation of graph completion, comparing the performance of our model, based on the updated NVD after applying \fixvw, and that of~\cite{zpshitops}, based on the original NVD.}
\label{tab:closedworldretrained} 
\begin{center}
    \begin{tabular}{|c||c|c|c|c|}
    \hline
   \textbf{CVE-CWE} & \textbf{Original~\cite{zpshitops}}  & \textbf{Fixed (Top-1)} & \textbf{Fixed (Top-2)} & \textbf{Fixed (Top-3)}\\
       \hline
        MRR & 0.445  & 0.452  & 0.662 & \textbf{0.731}\\
        MR & 13 &  12.164 & 6.455  & \textbf{5.177} \\
        Hits@20 & 0.840 & 0.855  & 0.931  & \textbf{0.943}\\
        Hits@10 & 0.731  & 0.758  & 0.878 &\textbf{ 0.903}\\
        Hits@3 & 0.504 & 0.529  & 0.757 & \textbf{0.809}\\
        Hits@1 & 0.309 & 0.303  & 0.532 & \textbf{0.627} \\
    \hline
    \end{tabular}
\end{center}
\end{table}

\subsection{Preventing Exploits}

Although accurate CWE mapping is not a direct prevention mechanism, it allows for earlier identification of the root causes of a vulnerability, which in turn facilitate patching. In this section, we focus on evaluating \fixvw\ for for exploited CVEs. We extract exploit data from the Known Exploited Vulnerabilities (KEVs) Catalog~\cite{cisa-kev-catalog} and the Exploit DB databases~\cite{exploitdb}.  The KEV list is maintained by Cybersecurity and Infrastructure Security Agency (CISA) and comprise vulnerabilities that are actively exploited in the wild, while the Exploit DB comprises a larger set of vulnerabilities with known exploits. 
Our combined dataset contains 21,468 exploits of which 9.7\% come from the KEV list and 90.3\% come from the Exploit DB list (we only consider exploits which have been independently verified). Figure~\ref{fig:cwe-24-dist} reveals that as of December 2024, 55.4\% of CVE are mapped to an invalid CWE or do not have a mapping. When looking at exploited CVEs specifically, this percentage increases to 62.55\% of CVEs that are mapped to invalid CWEs or do not have a mapping at all.

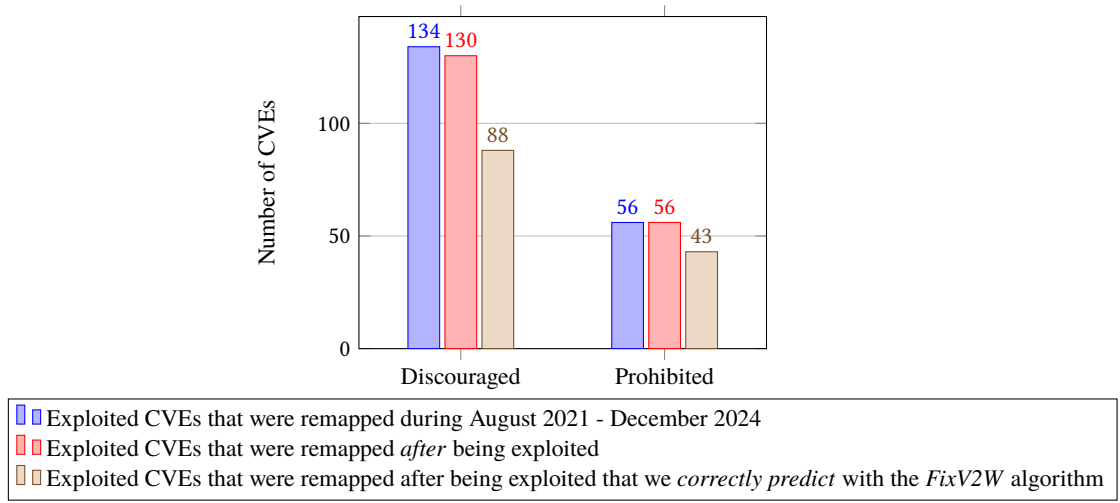
\begin{figure}[]
    \begin{tikzpicture}
    \pgfplotsset{width = 7cm, height = 6cm}
    \begin{axis}[
        ybar,
        bar width=12pt,
        symbolic x coords={Discouraged, Prohibited},
        xtick=data,
        ymin=0,
        ymajorgrids,
        ylabel={Number of CVEs},
        legend style={at={(0.5,-0.15)},
        anchor=north},
        legend cell align={left},
        nodes near coords,
        enlarge x limits=0.5
    ]
    \addplot coordinates {(Discouraged, 134) (Prohibited, 56)}; 
    \addplot coordinates {(Discouraged, 130) (Prohibited, 56)}; 
    \addplot coordinates {(Discouraged, 88) (Prohibited, 43)}; 
    
    \legend{Exploited CVEs that were remapped during August 2021 - December 2024, Exploited CVEs that were remapped \textit{after} being exploited, Exploited CVEs that were remapped after being exploited that we \textit{correctly predict} with the \fixvw\ algorithm}
    \end{axis}
    \end{tikzpicture}
    \centering
    \caption{KEV analysis August 2021 - December 2024}
    \label{fig:kev-data}
\end{figure}

To further motivate the importance of correct root cause mappings in the NVD, we analyzed the volume of exploits that happen after CWE remaps that we are able to correctly predict in 
Figure \ref{fig:kev-data}. 
It is seen that there are a total of 190 CVEs in the KEV list or Exploit DB that were exploited between August 2021 and December 2024 and were originally mapped to a Discouraged or Prohibited CWE. Out of these, 134 CVEs were mapped to a Discouraged CWE and 56 CVEs were mapped to a Prohibited CWE. \fixvw\ correctly predicts the root cause of a total of 131 CVE in the KEV list or in Exploit DB, which were not updated until after they were exploited. Keep in mind that \fixvw\ is trained on 2021 data and therefore would be able to correctly predict the root cause of these vulnerabilities.

\begin{table}[t]
\caption{Examples of Prohibited CVE-CWE mappings that were remapped after the exploit.}
\label{tab:kev-proh}
\begin{tabular}{|c|c|c|c|c|p{4cm}|}
\hline
\textbf{CVE} & \textbf{Exploit Date} & \textbf{Remap Date} & \textbf{Old CWE} & \textbf{New CWE} & \textbf{Description} \\
\hline
CVE-2011-1823 & 2022-09-08 & 2024-06-28 & CWE-189 & CWE-190 & Android OS Privilege Escalation \\
\hline
CVE-2013-0625 & 2022-03-07 & 2024-07-16 & CWE-255 & CWE-287 & Adobe ColdFusion Authentication Bypass  \\
\hline
CVE-2013-3993 & 2022-05-25 & 2024-06-28 & CWE-264 & CWE-22 & IBM InfoSphere BigInsights Invalid Input  \\
\hline
CVE-2015-1770 & 2022-03-28 & 2024-07-09 & CWE-19 & CWE-824 & Microsoft Office Uninitialized Memory Use  \\
\hline
CVE-2013-3897 & 2022-03-03 & 2024-07-16 & CWE-399 & CWE-416 & Microsoft Internet Explorer Use-After-Free  \\
\hline
\end{tabular}
\end{table}

\begin{table}[]
\center
\caption{Examples of Discouraged CVE-CWE mappings that were remapped after the exploit.}
\label{tab:kev-disc}
\begin{tabular}{|c|c|c|c|c|p{4cm}|}
\hline
\textbf{CVE} & \textbf{Exploit Date} & \textbf{Remap Date} & \textbf{Old CWE} & \textbf{New CWE} & \textbf{Description} \\
\hline
CVE-2007-5659 & 2022-06-08 & 2024-06-28 & CWE-119 & CWE-120 & Adobe Acrobat and Reader Buffer Overflow \\
\hline
CVE-2020-8644  & 2021-03-11 &	2022-07-12 &	CWE-74 & CWE-94	& PlaySMS Server-Side Template Injection \\
\hline
CVE-2018-13382 & 2022-01-10 & 2024-07-24 & CWE-285 & CWE-863 & Fortinet FortiOS and FortiProxy Improper Authorization \\
\hline
CVE-2012-0754 & 2022-06-08 & 2023-01-30 & CWE-119 & CWE-787 & Adobe Flash Player Memory Corruption \\
\hline
CVE-2014-0160 & 2022-05-04 & 2023-02-10 & CWE-119 & CWE-125 & OpenSSL Information Disclosure \\
\hline
\end{tabular}
\end{table}

Tables~\ref{tab:kev-proh} and~\ref{tab:kev-disc} show a total of 10 example known exploited vulnerabilities, that were exploited before the root cause was updated. The affected products include widely-used products such as Microsoft, Android, and Linux. Knowing the root cause and properly addressing could have possibly prevented these exploits from happening.

\begin{figure} []
\begin{minipage}{0.48\textwidth}
    
\begin{tikzpicture}
\begin{axis}[
    ybar stacked,
    bar width=12pt,
    width=\textwidth,
    xlabel={Rank},
    ylabel={Frequency},
    ymax = 16,
    xtick=data,
    xticklabels from table={\prohexp}{Rank},
    ymin=0,
    legend style={at={(0.5,-0.25)}, anchor=north, legend columns=3},
    title={Prohibited - Exploits}
]

\addplot+[draw = blue,line width = .2mm,fill=cyan] table[x expr=\coordindex, y=Exact] {\prohexp};
\addplot+[draw = red,line width = .2mm,fill = pink] table[x expr=\coordindex, y=Fine] {\prohexp};
\addplot+[draw = teal,line width = .2mm, fill=lime, nodes near coords, 
        point meta=y,
        nodes near coords align={anchor=south}] table[x expr=\coordindex, y=Coarse] {\prohexp};


\end{axis}
\end{tikzpicture}
\end{minipage}
\hfill
\begin{minipage}{0.48\textwidth}
\begin{tikzpicture}
\begin{axis}[
    ybar stacked,
    bar width=12pt,
    width=\textwidth,
    xlabel={Rank},
    ylabel={Frequency},
    ymax = 55,
    xtick=data,
    xticklabels from table={\discexp}{Rank},
    ymin=0,
    legend style={at={(0.5,-0.25)}, anchor=north, legend columns=3},
    title={Discouraged - Exploits}
]

\addplot+[draw = blue,line width = .2mm,fill=cyan] table[x expr=\coordindex, y=Exact] {\discexp};
\addplot+[draw = red,line width = .2mm,fill = pink] table[x expr=\coordindex, y=Fine] {\discexp};
\addplot+[draw = teal,line width = .2mm, fill=lime, nodes near coords, 
        point meta=y,
        nodes near coords align={anchor=south}] table[x expr=\coordindex, y=Coarse] {\discexp};

\end{axis}
\end{tikzpicture}
\end{minipage}
\vspace{0.6mm}
\begin{center}
\begin{tikzpicture}
\node[draw, inner sep=2pt] {
    \begin{tikzpicture}
\matrix [matrix of nodes, nodes={anchor=west}, column sep=2pt] {
    \node[draw=blue, fill=cyan, line width=.25mm, minimum width=10pt, minimum height=10pt]{}; & \node[]{Exact Match}; &
    \node[draw=red, fill=pink, line width=.25mm, minimum width=10pt, minimum height=10pt]{}; & \node[]{Fine Grain}; &
    \node[draw=teal, fill=lime, line width=.25mm, minimum width=10pt, minimum height=10pt]{}; & \node[]{Coarse Grain }; \\
};
\end{tikzpicture}
 };
\end{tikzpicture}
\end{center}
\caption{Breakdown of exact, fine, and coarse grain matches of exploited CVEs whose subsequent remaps are predicted correctly by \fixvw.}
\label{fig:exp-breakdown}
\end{figure}
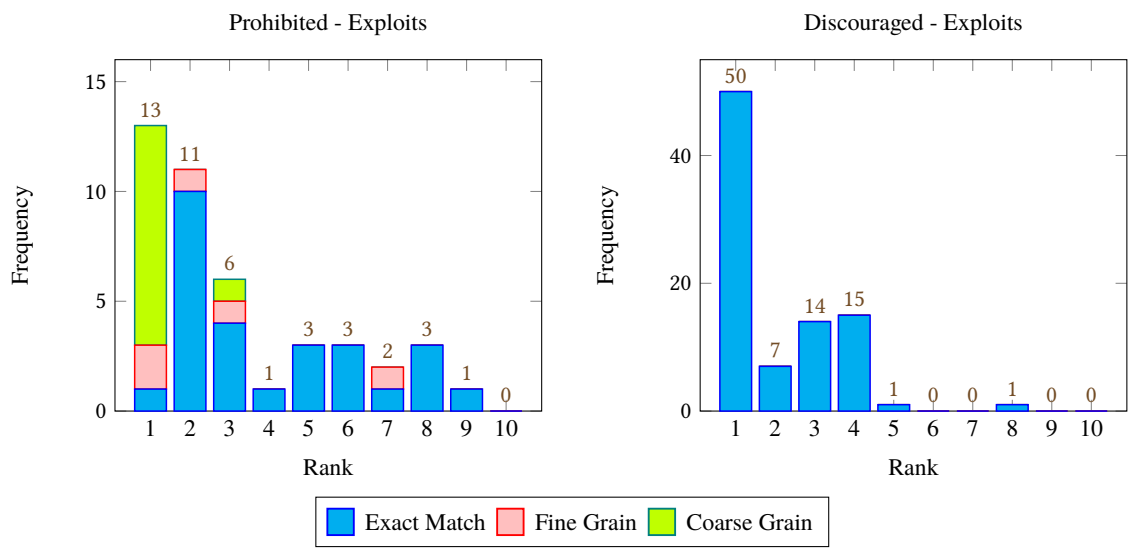

We look further into the distribution of CVE-CWE mappings that we are able to predict correctly. Figure \ref{fig:exp-breakdown} shows the distribution of CWE mappings that we are able to predict correctly as an exact, fine, or coarse match before the vulnerability was exploited. Notably, for the Discouraged set, all the exploits whose vulnerabilities' CWE remaps we predict correctly are exact matches, with all but one CVE ranking among the Top-5, indicating that \fixvw would have identified these remappings effectively before the CVE was exploited.

\pgfplotstableread{
Rank  Exact  Fine  Coarse  Overall
1 1 2 10 13
2 10 1 0 11
3 4 1 1 6
4 1 0 0 1
5 3 0 0 3
6 3 0 0 3
7 1 1 0 2
8 3 0 0 3
9 1 0 0 1
10 0 0 0 0
}\prohexp

\pgfplotstableread{
Rank  Exact  Fine  Coarse  Overall
1 50 0 0 50
2 7 0 0 7
3 14 0 0 14
4 15 0 0 15
5 1 0 0 1
6 0 0 0 0
7 0 0 0 0
8 1 0 0 1
9 0 0 0 0
10 0 0 0 0
}\discexp

\section{Discussion}\label{sec:discussion}


\subsection{Understanding Incorrect Predictions}\label{sec:incorrect}

For both the Prohibited and Discouraged CWE test sets, we find that in about 35-40\% of the cases, \fixvw\ fail to return a correct answer ranked within the Top-10 entries. We next investigate possible reasons that affect the model's performance. 
We first consider the Prohibited CWE test set.
It turns out that 61 entries correspond to CVEs that were previously mapped to CWE-16 or CWE-264, both of which do not have any members. Since they do not have any relationship in the Knowledge Graph ontology, it is difficult to find associations including with the \emph{FNN} method. Indeed, the \emph{FNN} method performs best with an ontology where meaningful relationships are established between the nodes, and CWE-16 and CWE-264 have few additional associations. 
Therefore \fixvw\ fails to find good replacements for CVEs that were previously mapped to these two CWE. In such cases, employing semantic models may be unavoidable. Unlike what is stated in our longitudinal analysis, where the rest of the CWE categories and members often remap to members, these two CWE are exceptional cases. Therefore, these CWE can be excluded from our test set or we can conclude that the method only works effectively if the initial CWE has meaningful mappings across the CWE hierarchy. Excluding these CWE from our test set, overall coverage of the test set goes up from 63\% to 75\% and exact match rate goes up from 40\% to 50\%. We reported the numbers without excluding these two in Section ~\ref{sec:eval} to not overfit our model and shrink our test set.  

For other incorrect predictions, we found that the new mapping is generally unrelated to the old CWE mapping, which means that the old mapping was highly 
inaccurate. For example, there are seven remaps from the old \emph{CWE-189: Numeric Errors} category that the model cannot predict. The new mappings in the NVD are as follows:
\begin{itemize}
    \item CVE-2003-2564 - CWE-776: Improper Restriction of Recursive Entity References in DTDs ('XML Entity Expansion')
    \item CVE-2007-3798 - CWE-252: Unchecked Return Value
    \item CVE-2008-0062 - CWE-665: Improper Initialization (Discouraged)
    \item CVE-2008-2108 - CWE-331: Insufficient Entropy
    \item CVE-2008-3612 - CWE-330: Use of Insufficiently Random Values
    \item CVE-2009-1890 - CWE-400: Uncontrolled Resource Consumption (Discouraged)
    \item CVE-2010-0128 - CWE-787: Out-of-bounds Write
\end{itemize}

None of these seven remaps are related to the ``Numeric Errors'' category, or any members of the category. In such cases, since the new CWEs are too far away in the ontology, the model does not recognize these as possible replacements to CWE-189.

Next, consider Discouraged mappings. We find that for about 40\% of the remaps, the new CWE mapping in a completely different branch than the old one. This cannot be avoided with our model. Moreover, if the new CWE is not a widely used weakness, even  candidates belonging to the CWE-1003 view and the Top-25 CWE view do not perform well. 

Yet, upon manually analyzing the new CVE-CWE mappings of these vulnerabilities, we found that some of the CVEs involved in these mappings are missing CWE mappings which our model actually predicts. For example, CVE-2007-3798 is caused by \emph{CWE-252: Unchecked Return Value}, which leads to \emph{CWE-190: Integer Overflow or Wraparound}, but the latter is not listed in the NVD. Our model predicts CWE-190 at rank 2. Similarly, CVE-2010-0128, caused by a signed integer error, should also be mapped to \emph{CWE-681: Incorrect Conversion between Numeric Types}, which is not listed in the NVD, while our model predicts it at rank 8. Hence, a reason our model does not always perform well is that our validation methodology assume that remaps from the NVD are correct and complete (in the absence of any better alternative), which is not always the case. 

\pgfplotstableread{
CWE  Correct  Incorrect 
CWE-16 3 4
CWE-189 65 6
CWE-19 1 0
CWE-254 3 3
CWE-255 10 7
CWE-264 25 57
CWE-275 1 0
CWE-310 31 20
CWE-320 1 5
CWE-399 76 20

}\Prohibitedbef

\pgfplotstableread{
CWE  Correct  Incorrect 
CWE-16 3 4
CWE-189 65 6
CWE-19 1 0
CWE-254 3 3
CWE-255 10 7
CWE-264 36 46
CWE-275 1 0
CWE-310 31 20
CWE-320 1 5
CWE-399 76 20

}\Prohibitedaft

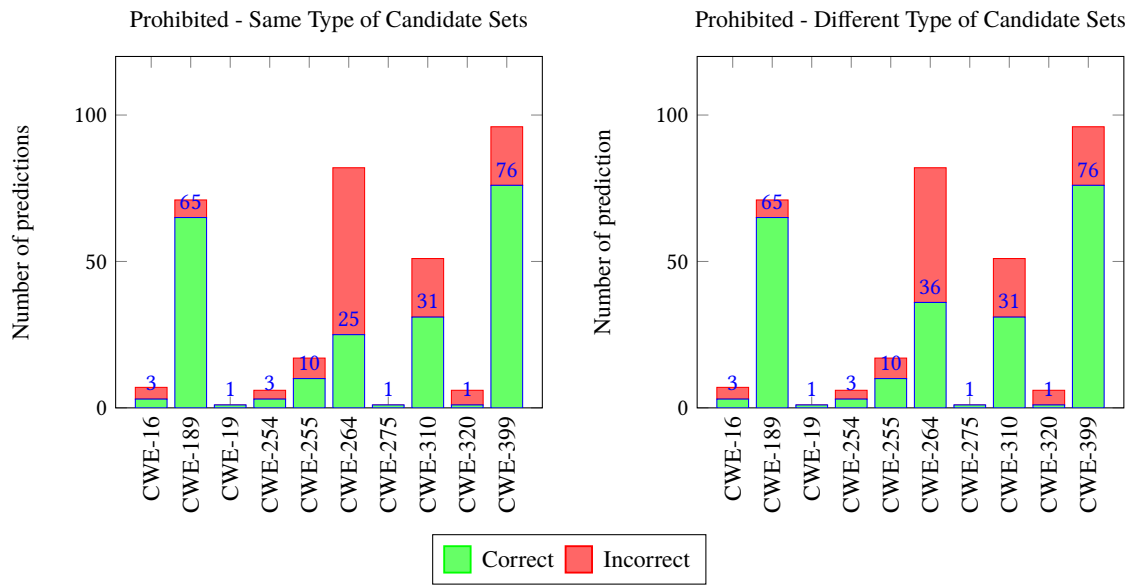
\begin{figure}[!t]
 
    \begin{minipage}{0.48\textwidth}
    \centering
    \begin{tikzpicture}
    \begin{axis}[
    ybar stacked,
    bar width=12pt,
    width=\textwidth,
    ylabel={Number of predictions},
    ymax = 120,
    xtick=data,
    xticklabels from table={\Prohibitedbef}{CWE},
    xticklabel style = {rotate=90, anchor=east},
    ymin=0,
    title={Prohibited - Same Type of Candidate Sets}
    ]

    \addplot+[fill=green!60, nodes near coords, 
        point meta=y,
        nodes near coords align={anchor=south}] table[x expr=\coordindex, y=Correct] {\Prohibitedbef};
    \addplot+[fill=red!60] table[x expr=\coordindex, y=Incorrect] {\Prohibitedbef};

    \end{axis}
    \end{tikzpicture}
    \end{minipage}
    \hfill
    \begin{minipage}{0.48\textwidth}
    \centering
    \begin{tikzpicture}
    \begin{axis}[
    ybar stacked,
    bar width=12pt,
    width=\textwidth,
    ylabel={Number of prediction},
    ymax = 120,
    xtick=data,
    xticklabels from table={\Prohibitedaft}{CWE},
    xticklabel style = {rotate=90, anchor=east},
    ymin=0,
    title={Prohibited - Different Type of Candidate Sets}
    ]
    \addplot+[fill=green!60, nodes near coords, 
        point meta=y,
        nodes near coords align={anchor=south}] table[x expr=\coordindex, y=Correct] {\Prohibitedaft};
    \addplot+[fill=red!60] table[x expr=\coordindex, y=Incorrect] {\Prohibitedaft};


    \end{axis}
    \end{tikzpicture}
    \end{minipage}
    \vspace{0.6mm}
    \begin{center}
    \begin{tikzpicture}
    \node[draw, inner sep=2pt] {
    \begin{tikzpicture}
    \matrix [matrix of nodes, nodes={anchor=west}, column sep=2pt] {
        \node[draw=green, fill=green!60, line width=.25mm, minimum width=10pt, minimum height=10pt] {}; & \node[] {Correct}; &
        \node[draw=red, fill=red!60, line width=.25mm, minimum width=10pt, minimum height=10pt] {}; & \node[] {Incorrect}; \\
        };
        \end{tikzpicture}
    };
    \end{tikzpicture}
    \end{center}
    
    \caption{Prohibited - Comparison between using the same and different type of candidate sets for each CWE. Counts shown correspond to the number of correct remappings.}
    \label{fig:proh-per-bef}
\end{figure}

\subsection{Using Different Types of Candidate Sets For Each CWE}
\begin{figure}[!t]
\centering
\begin{minipage}{0.48\textwidth}
\begin{tikzpicture}
\begin{axis}[
    ybar stacked,
    bar width=12pt,
    width=\textwidth,
    ylabel={Number of predictions},
    ymax = 600,
    xtick=data,
    xticklabels from table={\Discouragedbef}{CWE},
    xticklabel style = {rotate=90, anchor=east},
    ymin=0,
    title={Discouraged - Same Type of Candidate Sets}
]

\addplot+[fill=green!60,
nodes near coords, 
        point meta=y,
        nodes near coords align={anchor=south}] table[x expr=\coordindex, y=Correct] {\Discouragedbef};
\addplot+[fill=red!60] table[x expr=\coordindex, y=Incorrect] {\Discouragedbef};

\end{axis}
\end{tikzpicture}
\end{minipage}
\hfill
\begin{minipage}{0.48\textwidth}
\centering
\begin{tikzpicture}
\begin{axis}[
    ybar stacked,
    bar width=12pt,
    width=\textwidth,
    ylabel={Number of prediction},
    ymax = 600,
    xtick=data,
    xticklabels from table={\Discouragedaft}{CWE},
    xticklabel style = {rotate=90, anchor=east},
    ymin=0,
    title={Discouraged - Different Type of Candidate Sets}
]

\addplot+[fill=green!60, nodes near coords, 
        point meta=y,
        nodes near coords align={anchor=south}] table[x expr=\coordindex, y=Correct] {\Discouragedaft};
\addplot+[fill=red!60] table[x expr=\coordindex, y=Incorrect] {\Discouragedaft};


\end{axis}
\end{tikzpicture}
\end{minipage}
\vspace{0.6mm}
    \begin{center}
    \begin{tikzpicture}
    \node[draw, inner sep=2pt] {
    \begin{tikzpicture}
    \matrix [matrix of nodes, nodes={anchor=west}, column sep=2pt] {
        \node[draw=green, fill=green!60, line width=.25mm, minimum width=10pt, minimum height=10pt] {}; & \node[] {Correct}; &
        \node[draw=red, fill=red!60, line width=.25mm, minimum width=10pt, minimum height=10pt] {}; & \node[] {Incorrect}; \\
        };
        \end{tikzpicture}
    };
    \end{tikzpicture}
    \end{center}
\caption{Discouraged - Comparison between using the same and different type of candidate sets for each CWE. Some CWE not shown in Discouraged set, where no change is observed. Counts shown correspond to the number of correct remappings.}
\label{fig:disc-per-bef}
\end{figure}
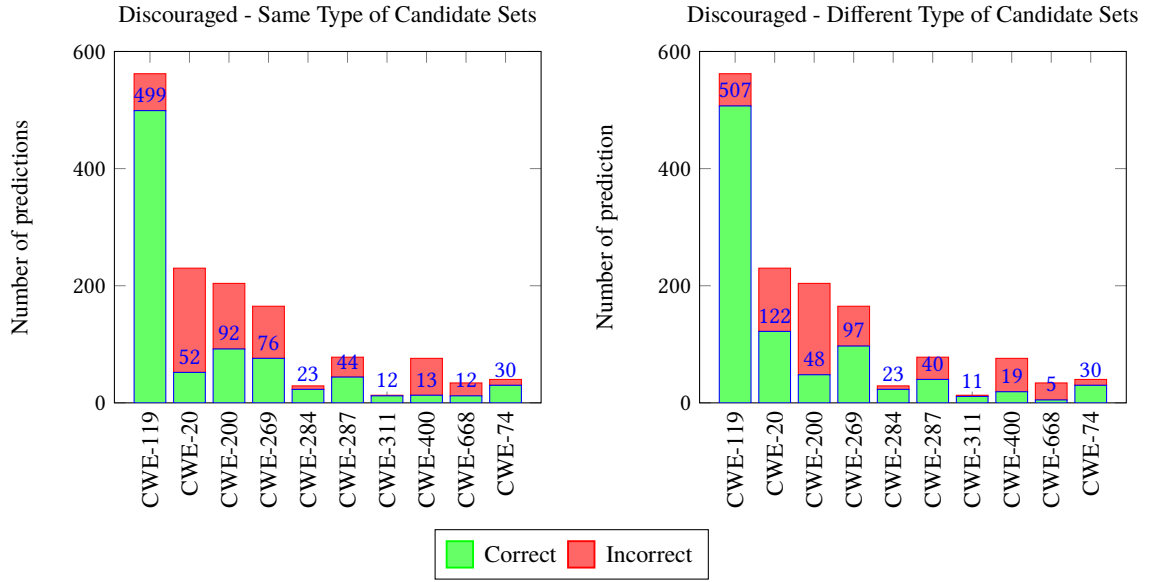

Instead of using the same type of candidate sets for all CWEs (e.g., Top-25, Descendants, etc.), we experimented with tailoring the candidate set for each CWE. We used remaps from before August 2021 to make predictions on remaps after August 2021. We used the set that the majority of CVEs before August 2021 were remapped to to determine the type of candidate set we query after 2021. The different types of candidate sets we tested were Top-25, Descendants, Member, and Neighbor for each CWE. For the Prohibited set, we compared the performance of this tailored algorithm to the \textit{Member} algorithm in Figure \ref{fig:proh-per-bef}. For the Discouraged set, we compared the performance of this tailored algorithm to the \textit{Family} algorithm in Figure \ref{fig:disc-per-bef}. Surprisingly, some CWEs such as CWE-20 perform best when queried with baseline candidate sets (i.e. CWE Top-25), but for other CWE such as CWE-200, using this algorithm actually has an inverse effect on model performance. 

\pgfplotstableread{
CWE  Correct  Incorrect 
CWE-119 499 63
CWE-20 52 178
CWE-200 92 112
CWE-269 76 89
CWE-284 23 6
CWE-287 44 34
CWE-311 12 1
CWE-400 13 63
CWE-668 12 22
CWE-74 30 10
}\Discouragedbef

\pgfplotstableread{
CWE  Correct  Incorrect 
CWE-119 507 55
CWE-20 122 108
CWE-200 48 156
CWE-269 97 68
CWE-284 23 6
CWE-287 40 38
CWE-311 11 2
CWE-400 19 57
CWE-668 5 29
CWE-74 30 10

}\Discouragedaft

\subsection{FixV2W vs NVD: Which is more accurate?}
\label{sec:more-accurate}

Upon the completion of our predictions and evaluation, we manually analyzed some of our predictions to determine how accurate they are.
As discussed in Subsection \ref{sec:incorrect}, we found out that \fixvw\ returns CWEs that are missing from the NVD data, especially when CVEs are complex and require multiple CWE mappings. Indeed, often the NVD provides only one mapping, while in practice others CWEs also contribute to a vulnerability. 
Apart from this, we discovered instances where \fixvw\ appears to provide a more precise mapping than the NVD, as shown next.

For instance, CVE-2016-5042 is caused by a crafted input (malformed DWARF section) that leads to an infinite loop and a crash. The NVD maps this CVE to \emph{CWE-835: Loop with Unreachable Exit Condition}. Yet the primary root cause is CWE-20, Improper Input Validation, which leads to the malformed section being accepted as valid input. \fixvw\ predicts CWE-20 at the first rank.

Similarly, CVE-2014-1943 is caused by a crafted indirect offset value in the magic of a file, and that allows an attacker to cause denial of service (infinite recursion, CPU consumption, and crash). The NVD maps this CVE to \emph{CWE-755: Improper Handling of Exceptional Conditions}. Upon careful analysis, it appears that this vulnerability is caused by input parsing logic allowing recursive evaluation without termination conditions. This directly maps to CWE-400 (Uncontrolled Resource Consumption), CWE-770 (Allocation of Resources Without Limits) or CWE-772 (Missing Release of Resource After Effective Lifetime); all of which are predicted by \fixvw\ in ranks 8,9 and 4, respectively. The primary issue, the program not checking the input to ensure no recursive entities are used (CWE-20) is also predicted by our model at the first rank. 

In total, we detected with the help of ChatGPT-4o (verified manually afterwards), that 11 of the 123 ``incorrect'' predictions in the Prohibited CWE set are in fact correct CWEs that are not listed by the NVD. These correctly predicted CWE are missed in the evaluation pipeline of \fixvw, since the NVD fails to provide them or a related CWE in the vulnerability metadata.

\subsection{Shortcoming of semantic models}  Accurate correction of invalid mappings is far from a straightforward task, even for semantics-based models that are trained  on large amount of data. These models, while powerful, require large, well-labeled training datasets that include textual descriptions for both CVEs and CWEs. LLMs struggle with consistency and lack deterministic responses~\cite{atil2025nondeterminis}, rendering them unreliable as a standalone approach to CVE classification~\cite{marchiori2025llmscve}. For example, consider CVE-2021-21589  that lacks a CWE mapping. Because the CVE description mentions ``adversary can use this to escalate privileges'', ChatGPT-4o suggests \emph{CWE-269: Improper Privilege Management}. Yet, the actual issue is the program not exiting on failed initialization i.e. \emph{CWE-455: Non-exit on Failed Initialization} which is the child of \emph{CWE-665: Improper Initialization}. Such issues are hard to avoid using semantic processing methods, such as LLM or NLP.  We argue that using the natural connectivity of CVEs, CWEs and CPEs through an ontology representation of the NVD data can help inform the predictions of semantic models and reduce computational overhead. In this work, we made the first effort to correct CVE-CWE mappings to improve database quality, and increase model accuracy of those what train on NVD data. Unlike large language models, \fixvw\ avoids keyword overfitting and provides interpretable, reproducible predictions that align with ontology structure. 

\subsection{Limitations and Threats to Validity}\label{sec:limit}

\subsubsection{Data Accuracy}
For both our predictions and validations, we rely on the National Vulnerability Database (NVD). Our analysis begins with the understanding that the NVD contains known inaccuracies in CWE mappings.
In our current study, the updated version of the NVD serves as the validation set under the assumption that its updated mappings represent the most accurate classification available. However, errors and inconsistencies persist even in recent updates, as discussed earlier. This directly affects the reported performance of our model, as validation accuracy is bounded by the correctness of these reference labels. For instance, when a CVE is mapped by the NVD to a generic (Discouraged) CWE despite the availability of more specific alternatives, our model may be penalized for offering the correct, but unlisted, alternative. Similarly, the model is penalized for offering correct, but unlisted root cause weaknesses, as discussed in Subsection~\ref{sec:incorrect}. 


\subsubsection{Results Evaluation}
Our results include both fine-grained and coarse-grained CWE predictions, each of which contributes to reducing the manual burden of remapping vulnerabilities. Nonetheless, our evaluation strategy currently assumes that the most recent CWE mapping provided by the NVD or its associated CNA is not only correct but also complete. As explained earlier, this is not always the case. Indeed, certain vulnerabilities require multiple CWE mappings to fully capture the root causes and contributing factors. The NVD, however, often lists only a single CWE per CVE, even when the vulnerability spans multiple weakness types. This simplification leads to under-specified mappings and impacts both model training and evaluation, as the complexity of some vulnerabilities are not reflected in the data. 


\section{Conclusion}\label{sec:conclusion}

This paper presents \fixvw, a knowledge graph–based framework designed to correct invalid CVE–CWE mappings in the NVD. Through a longitudinal study of mapping updates between 2016 and 2024, we uncovered that more than half of all CVEs are currently assigned to invalid CWE categories, with 84\% of subsequent remaps occurring within just one to two hops in the CWE hierarchy. Leveraging these insights, we design \fixvw, a methodology that systematically identifies more accurate weaknesses. Remarkably, FixV2W provides correct matches (among the Top-10 results) for 69\% of exploited vulnerabilities that had invalid CWEs before they were exploited. Moreover, FixV2W demonstrates multi-label classification ability, shown with the improvements to the original model after retraining with top 2 and top 3 matches. Strikingly, replacing invalid mappings with multiple mappings predicted by FixV2W improved the model's ability to predict unseen CVE-CWE mappings significantly, and provides more meaningful root cause mappings than a single label classification. 

We compare different candidate sets to achieve best results, and our tests show that for Discouraged CWE correction, using CWEs closely related in the hierarchy (family) performs best, while for Prohibited CWE correction, using CWEs that are members of the same category or view of the original CWE yields best results. Even further, excluding a few outliers that do not fit into the longitudinal analysis claims (no members, few associations) proves that the model performs best if the old CWE mapping has strong, meaningful connections in the ontology. 
Our results demonstrate that ontology-informed graph embeddings provide a practical and deterministic alternative to purely semantic or LLM-based approaches, avoiding hallucinations and nondeterministic responses while remaining computationally lightweight.

Beyond improving CWE mapping accuracy, our findings highlight broader implications for vulnerability management: higher-quality root cause data enables better future predictions and overall improved ML model performance throughout the cybersecurity ecosystem. While our model misses correct CWEs in the predictions, some original invalid mappings are substantially misaligned with the true weakness, limiting purely structural approaches, such as ours. As explained in Section~\ref{sec:limit}, NVD does not always provide the most accurate CWE remap for a CVE, and sometimes \fixvw\ outperforms the ``ground truth''.

A promising direction for future work is to integrate \fixvw\ with semantic models, particularly large language models (LLMs), to further improve and validate our model's performance in correcting CWE mappings of CVE. Such an approach could combine \fixvw’s deterministic, ontology-grounded predictions with the contextual reasoning strengths of LLMs, providing both precision and adaptability as new weaknesses emerge. We believe this integration will pave the way toward the next generation of automated, trustworthy, and scalable vulnerability intelligence systems.

\begin{acks}
This work was supported in part by the Boston University Red Hat Collaboratory (awards numbers~2024-01-RH03 and 2025-01-RH05). The authors thank David Sastre Medina, Jonah Gluck and Howell Xia for fruitful discussions. 
\end{acks}

\bibliographystyle{ACM-Reference-Format}
\bibliography{sample-base}


\end{document}